\documentclass[
preprint,
 aps,
prb,
]{revtex4-2}

\usepackage[encapsulated]{CJK}
\usepackage{multirow}
\usepackage{amsmath}
\usepackage{mathtools}
\usepackage{amssymb}
\usepackage{mathrsfs}
\usepackage{graphics}
\usepackage{graphicx}
\usepackage{rotating}
\usepackage{notoccite}
\usepackage{bbold}
\usepackage{bm}
\usepackage{hyperref}
\DeclareMathOperator{\Tr}{Tr}
\begin{document}

\title{Symmetry constraints on topological invariants and irreducible band representations}

\author{Jing Zhang}
\affiliation{Department of Physics and Institute of Advanced Studies, Tsinghua University, Beijing 100084, PRC and Blackett Laboratory, Department of Physics, Imperial College London, Prince Consort Road, London SW7 2BW, UK}

\date{\today}

\begin{abstract}
Elementary band representation (EBR) is considered the building block in topological quantum chemistry and fundamental concept in symmetry indicator methods. One of the hypophysis is that a fully occupied EBR has zero Berry-Wilczek-Zee (BWZ) phase $\phi_B$ and those occupied corresponds to trivial topology. Associated with it are the concepts of atomic limit and equivalence between band representations. In this manuscript, an explicit link between the BWZ phase of physical dispersion surfaces and that of its EBR or irreducible band representation (IBR) bases is established. When gapped system occurs under the tight binding model, the relation between the BWZ phase of a set of connected bands and its band representation bases only persist if the later are IBRs. Thus the BWZ phase can be evaluated in terms of the IBR bases. Equivalent segments of path integral of BWZ connection with respect to IBRs are established as representation of the space group and selection rule for corresponding BWZ phase established where possible. The occurrence of IBRs is rooted in real space symmetry but dependent on dynamic interaction and band topology. Three gapped systems in honeycomb lattices with layer group 80(L) are discussed. Two spin-less cases are shown to be topologically trivial (symmetry forbidden $\phi_{\rm B}$) where as the selection rule cannot be developed for the spin-full ${\rm p_z}$ orbital as in graphene. Two necessary conditions for topologically trivial phase are established, namely 1. Connected bands having the same closed set of IBR bases for all ${\rm\bf k}$ in Brillouin zone and, 2. The reduced tensor element for the path integral of BWZ connection for such bases is symmetry forbidden due to contractable close loop having zero BWZ phase. Thus the IBRs are the building block of topologically trivial phase of connected bands and symmetry constraint on BWZ phase are obtained through IBRs and selection rules via Wigner-Eckart theorem. A number of illustrated cases demonstrate that the basic hypothesis of symmetry indicator method is false. The analysis here advocate a paradigm shift from EBR to IBR as building block of topologically trivial phase in quantum chemistry.
\end{abstract}


\maketitle
\section{Introduction}
Symmetry protected topological state (SPTS) \cite{Wen_X_G:2004} in crystalline condensed mater are often characterised by Chern numbers, or Berry-Wilczek-Zee (BWZ) phase $\phi_{\rm B}$ \cite{Berry_M_V:1984, Wilczek_F:1984, Alexandradinata_A:2014} of the connected bands in the Brillouin Zone (BZ). The BZ, a compact, boundary-less (closed) real manifold, serves as the base manifold to the vector (Bloch) bundle. In one, two, and three dimensional crystals, the BZ correspond to a circle, 2-torus $T^2$, and 3-torus $T^3$ respectively (see Fig.\ref{fig:BZ} for an illustration of 2D case based on graphene). They are compact but not simply connected. A set of connected band forms the vector bundle with the BZ as the base manifold. The BWZ phase is then defined as cohomology of the BWZ connection:
\begin{equation}
\phi_B=-i\ln \det \left[W(\gamma)\right]=-i\Tr\oint_{\gamma}\left<\psi_m({\rm\bf k})\mid\nabla_{\rm\bf k}\mid\psi_n({\rm\bf k})\right>\cdot{\rm d \bf k}
\label{eqn:BWZphase}
\end{equation}
where ${\rm\bf k}$ is the wave-vector (element in the BZ manifold), $\gamma$ is any {\em un-contractable closed} path in the BZ,  $\left|\psi_n({\rm\bf k})\right>$ is the energy eigenstate (element of the vector bundle section) with $m, n$ as index to the connected physical bands and $W(\gamma)$ is the Wilson loop describing the evolution of multi-band eigenstates under adiabatic transport along $\gamma$\cite{Weinberg_S:2015, Bradlyn_B:2022}. In topological band theory, much of the investigation was focused on elementary band representations (EBR). Band representations (BR)\cite{Zak_J:1981A, Zak_J:1982A, Evarestov_R_A:1984A, Evarestov_R_A:1984B, Elcoro_L:2017}  were first proposed by Zak in the 1980's and EBR is considered building block of topological quantum chemistry\cite{Bradlyn_B:2017, Cano_J:2018B}. These studies are based on the hypothesis of zero $\phi_{\rm B}$ from EBRs and any physical connected bands compatible with composite EBRs (CBR). The classification of topological phases were also made using combinatorics based on compatibility relations\cite{Kruthoff_J:2017}. The connection between $\phi_{\rm B}$ of physical connected bands and composite EBR bases is not well identified and the role of symmetry is not explicitly known. The dynamically occurring irreducible band representations(IBR)\cite{Zhang_J:2021} and its roles are not recognised. Symmetry, as in many selection rules\cite{Eckart_C:1930}, can  forbid a physical quantities such as $\phi_{\rm B}$ from taking non-zero values. In other words, one must be able to establish selection rules which identify phases of connected bands that are symmetry guaranteed to be topologically trivial (symmetry forbidden $\phi_{\rm B}$). Absence of such selection rules means the phase is potentially (or is symmetry indicated as) non-trivial. 

In the accompanying manuscript\cite{Zhang_J:2021}, the concept of dynamically occurring IBR, arising from linear combination of EBR components on the Wyckoff position index, is introduced. An EBR occurring in a set of fully connected bands is an IBR itself. For the purpose of symmetry analysis, the {\em path integral of Berry-Wilczek-Zee connection} (PIBC) with respect to IBRs and corresponding $\phi_{\rm B}^{\rm IBR}$ are defined for the same path $\gamma$ in the BZ
\begin{equation}
\mathcal{BWZ}(\gamma)_{i,j}^{\rm IBR}=\int_{\gamma}\left<\phi_i({\rm\bf k})\mid\nabla_{\rm\bf k}\mid\phi_j({\rm\bf k})\right>\cdot{\rm d \bf k}, \;\;\;\;
\phi_B^{\rm IBR}=-i\Tr \mathcal{BWZ}(\gamma)_{i,j}^{\rm IBR}
\label{eqn:BWZphaseBR}
\end{equation}
where $\left|\phi_j({\rm\bf k})\right>$ are the Wannier function induced bases (EBRs) or dynamically occurring IBRs arising from linear combinations of EBR components for the tight binding (TB) model, and $i,j$ are the composite index of EBRs/IBRs instead of physical band index $m, n$ in Eq.\eqref{eqn:BWZphase}. We use $\gamma$ to indicate closed un-contractable path and $\lambda, \kappa$ refers to general path with start point ${\rm\bf k}_a$ and end point ${\rm\bf k}_b$ in the BZ. In this manuscript, the relation between $\phi_{\rm B}$ of physical bands and $\phi_{\rm B}^{\rm IBR}$ of EBR/IBR bases is established through the TB model. If an isolated connected physical bands have a full description in terms of the {\em same set of IBR bases independent of ${\rm\bf k}$}, $\phi_{\rm B}$ can be evaluated via $\phi_{\rm B}^{\rm IBR}$ with respect to the relevant EBRs/IBRs. This avoid the issue of degeneracy which may occur among connected physical bands in the BZ and potentially the SU(N) gauge group one needs to deal with in the path integral. Utilising the transformation properties of EBR/IBR and PIBC forming representation of space group, the constraint of space group symmetry on $\phi_{\rm B}$ is established through Wigner Eckart theorem. Analysis shows the symmetry does not guarantee trivial phases for all connected bands with IBR description. Hence the symmetry indicator method is not based on selection rule derivable from space groups.  For some space groups, selection rules on $\phi_{\rm B}^{\rm IBR}$ can be established through forbidden reduced tensor elements of PIBC. Only in these cases the connected bands are symmetry guaranteed topologically trivial. As the properties of the whole BZ is under investigation, it is necessary to adopt the full group method\cite{Bradley_C_J:2010, Birman_J:1966} which covers all equivalent paths generated by symmetry operation in the BZ. Examples are given for connected band in split EBRs which are topologically non-trivial (absence of selection rule), trivial (symmetry forbidden $\phi_{\rm B}$), and connected bands equivalent to composite IBRs which are topologically trivial (symmetry forbidden $\phi_{\rm B}$), all based on honeycomb lattice with sub-periodic layer(space) group $\mathcal{G}=80$(L). This includes the gapped phase of graphene, from which a simple conclusion of absence of selection rule can be drawn.

\section{General approach in symmetry analysis of Berry phase}
Impact of symmetry on physical observables, perhaps, sees its best manifestation in the Wigner-Eckart theorem\cite{Eckart_C:1930} or its finite group equivalent\cite{Koster_G:1958}. The theorem affords us selection rules (if the observable is symmetry forbidden) and relative strength of matrix elements, and can be applicable to the BWZ phase derived from PIBC (which give rise to the Wilson loop operator when exponentiated). The BWZ phase, as a topological invariant of a given connected bands/Bloch bundle in crystalline matter, is clearly invariant under any symmetry operation of the corresponding space group $\mathcal{G}$. It is the trace of closed uncontactable PIBC with respect to eigenstates in the connected band. It is difficult to consider PIBC with respect to physical bands as representation of space group due to the ${\rm\bf k}$ dependence of transformation properties of its constituent, and the existence of degeneracy. The latter requires consideration of SU(N) gauge group. However, the $\phi_{\rm B}$ with respect to eigenstates of physical bands can be calculated in terms of its BR bases if such bases are closed direct sum of IBRs (see Sec.\ref{sec:ChernEBR}). This gives rise to the first necessary condition for the existence of selection rule on $\phi_{\rm B}$. 

The first necessary condition facilitates the evaluation of $\phi_{\rm B}$ in term of closed complete set of IBR bases subject to maintenance of energy gap. In the context of BWZ connection, what form a representation of the space group are the {\em equivalent} PIBC with respect to the IBRs generated by the action of the space group. This is possible because such bases (IBRs) have well defined transformation properties with no explicit ${\rm\bf k}$ dependance on the rotational part. Given a symmetry operation $\{R|{\rm\bf v+t}\}\in\mathcal{G}$ in real space ($R$ is an element of point group F of the crystal with associated non-lattice translation ${\rm\bf v}$ and $\rm\bf t$ is a lattice translation), it leads to the following transformation in the BZ/vector bundle, and EBRs/IBRs.
\begin{subequations}\label{eqn:sym_tfm}
	\begin{align}
		{\rm\bf k}&\rightarrow~{\rm\bf k}^\prime=R{\rm\bf k} \\
		\gamma&\rightarrow~\gamma^\prime  \\
		\left|\psi_m({\rm\bf k})\right>&\rightarrow~\left|\psi_n({\rm\bf k}^\prime)\right> \label{eqn:sym_tfm_c}\\ 
		\left|\phi_j({\rm\bf k})\right>&\rightarrow~\left|\phi_i({\rm\bf k}^\prime)\right> \\ 
		\nabla_{\rm\bf k} \{\bullet\} \cdot \mathrm{d}{\rm\bf k}&\rightarrow~\nabla_{\rm\bf k^\prime} \{\bullet\}\cdot \mathrm{d}{\rm\bf k}^\prime 
	\end{align}
\end{subequations}
The definition of the BWZ phase in Eq.\eqref{eqn:BWZphase} requires the equivalent paths to be closed and un-contractable. But any part of $\gamma$ can form representation of $\mathcal{G}$ and their sum may combine to form the {\em un-contractable or contractable closed loop}. As the equivalent paths must extend beyond the representation domain, it is necessary to consider the full group method\cite{Bradley_C_J:2010, Birman_J:1966} rather than discussion based on  group of ${\rm\bf k}$ within the representation domain. Since $\nabla_{\rm\bf k}$ and $\mathrm{d}{\rm\bf k}$ transform contragrediently to each other, together they transform identically between the equivalent paths (i.e. $\nabla_{\rm\bf k} \{\bullet\} \cdot \mathrm{d}{\rm\bf k}=\nabla_{\rm\bf k^\prime} \{\bullet\}\cdot \mathrm{d}{\rm\bf k}^\prime$).  Thus the integrand of Eq.\eqref{eqn:BWZphase} (and Eq.\eqref{eqn:BWZphaseBR}) transform according to the product of a pair of representations conjugate to each other due to those arising from $\left|\psi_n({\rm\bf k})\right>$ and $\left<\psi_m({\rm\bf k})\right|$ (the basis $\left|\phi_j({\rm\bf k})\right>$ and $\left<\phi_i({\rm\bf k})\right|$). The action of $\mathcal{G}$ then permute the PIBC among its equivalent paths. One can conclude that  at least one trivial representation of the space group $\mathcal{G}$ is contained in the representations formed by the equivalent CPIBC. 

In search for trivial insulator phase (zero $\phi_B$), one can apply the Winger-Eckart theorem to PBIC with respect to IBR bases and identify what constraints (selection rules) that may exist on the BWZ phase either through the traceless `angular dependent matrix' or zero `reduced tensor element'. This can then be passed on to the $\phi_B$ of physical bands if the first necessary condition is satisfied. It is important to recognise that segments of the CPIBC may form representations of $\mathcal{G}$, in addition to the complete closed un-contractable loops. 

When electron spin is considered in addition to its orbital motion, there is a distinction between sub-periodic layer group (2D symmetry embedded in 3D space) and plane group (2D symmetry). For example, the point group contained in the layer group of graphene is ${\rm D}_{6h}$ whereas that of the plane group is C$_{6v}$. Due to 2D nature of plane group, its point group is always a direct product involving subgroup of SO(2), {\em not} SO(3). Since, SU(2) is only the double cover of SO(3), spin is not relevant in the case of plane group. When spin degree of freedom is considered, it is only appropriate to consider the sub periodic layer group\cite{ITFCE} where the 2D symmetry is embedded in 3D space.

\section{BRs and IBRs}
\label{sec:BRs}
BR, first proposed by Zak in 1981, are Wannier functions induced representation in ${\rm\bf k}$ space. They are the bases of TB Hamiltonian as described by Slater and Koster\cite{Slater_J:1954}.  The band representation as defined by Zak is as follows
\begin{equation}
\Phi^{\rm\bf k}_{\bm{\tau}_\alpha, \mu, i}({\rm\bf r})= \left<{\rm\bf r}|{\rm\bf k},{\bm \tau}_\alpha,\mu,i\right>_{\rm Zak}=\Omega^{-1}\sum_{\ell}\exp\{i{\rm\bf k}\cdot {\rm\bf t}_\ell\}W^{\mu, i}_{\bm{\tau}_\alpha}({\rm\bf r}-({\rm\bf t}_\ell+\bm{\tau}_\alpha)) \label{eqn:Zak_basis}
\end{equation}
where $W^{\mu, i}_{\bm{\tau}_\alpha}({\rm\bf r}-({\rm\bf t}_\ell+\bm{\tau}_\alpha))$ is the Wannier function centred on the Wyckoff position ${\rm\bf t}_\ell+{\bm\tau}_\alpha$ and transform according to irreducible representation (irrep) $\mu$ of the site symmetry group $\tilde{\rm G}^{\bm{\tau}}$ of Wyckoff position ${\bm\tau}_\alpha$. $\ell$ index the primitive unit cells and $\alpha$ index the equivalent Wykoff positions. ${\bm\tau}_\alpha$, $\mu$, and $i$ make up the composite index $j$ of EBR used later in Eq.\eqref{eqn:unitary}.  This definition differs from the bases of Slater Koster formulation\cite{Slater_J:1954} of TB Hamiltonian
\begin{equation}
\phi_{\bm{\tau}_\alpha, \mu, i}^{\rm\bf k}({\rm\bf r})=\left<{\rm\bf r}|{\rm\bf k},{\bm \tau}_\alpha,\mu,i\right>_{\rm TB}=\Omega^{-1}\sum_{\ell}\exp\{i{\rm\bf k}\cdot ({\rm\bf t}_\ell+{\bm \tau}_\alpha)\}W^{\mu, i}_{\bm{\tau}_\alpha}({\rm\bf r}-({\rm\bf t}_\ell+\bm{\tau}_\alpha))
\label{eqn:TB_basis}
\end{equation}
by a ${\rm\bf k}$ dependent  gauge transformation:
\begin{equation}
\phi^{\rm\bf k}_{\bm{\tau}_\alpha,\mu, i}({\rm\bf r})=\exp(i{\rm\bf k}\cdot\bm{\tau}_\alpha) \Phi^{\rm\bf k}_{\bm{\tau}_\alpha, \mu, i}({\rm\bf r}).
\label{eqn:gauge_transform}
\end{equation}
Both are well defined for all ${\rm\bf k}\in$ BZ, but there are linear dependence between ${\rm\bf k}^\prime$ and ${\rm\bf k}$ on the surface of the first BZ if ${\rm\bf k}^\prime={\rm\bf k}+{\rm\bf g}$. Here $\rm\bf g$ is a reciprocal lattice vector. The Zak basis obeys $\left|{\rm\bf k}, {\bm \tau}_\alpha, \mu, i\right>_{Zak}=\left|{\rm\bf k}+{\rm\bf g}, {\bm \tau}_\alpha, \mu, i\right>_{Zak}$ whereas the TB basis requires additional phase factor $\exp(i{\rm\bf g}_R\cdot{\bm \tau}_\alpha)$ in the relevant representation matrices. Whilst both bases form representations of the space group at a given ${\rm\bf k}$, their transformation matrices are different. The transformation properties of the Zak basis are described in Ref.\onlinecite{Evarestov_R_A:1984A, Evarestov_R_A:1984B,Cano_J:2018B}. For the TB basis of the full vector space spanned by EBR\cite{Zhang_J:2021}
\begin{align}
&\Gamma_{\tilde{\rm G}^{\bm{\tau}}\uparrow}^{\mathcal{G}}(\{R|{\rm\bf v+t}\})_{{\rm\bf k}^\prime, \bm{\tau}_\alpha^\prime, \mu^\prime, i^\prime; {\rm\bf k}, \bm{\tau}_\alpha, \mu, i}=\left<{\rm\bf k}^\prime, \bm{\tau}_\alpha^\prime, \mu^\prime, i^\prime \mid \hat{S}(\{R|{\rm\bf v+t}\}\mid {\rm\bf k}, \bm{\tau}_\alpha, \mu, i\right>\notag \\
&\quad\quad=\exp\{-iR{\rm\bf k}\cdot({\rm\bf v+t})\}  D^{\rm\bf k}(R)_{{\rm\bf k}^\prime, {\rm\bf k}}\underbrace{\exp\{i{\rm\bf g}_R\cdot\bm{\tau}_\alpha^\prime\}\delta_{\mu^\prime,\mu}D^{\bm{\tau},\mu}(R)_{m^\prime, m}}_{\Gamma_\phi}.\label{eqn:EBR} 
\end{align}
where ${\rm\bf k},{\rm\bf k}^\prime\in\{\ast{\rm\bf k}\}$, $m^\prime, m$ are composite labels of $\tau_\alpha, \mu, i$,  and
\begin{align}
\begin{array}{ll}
	D^{\rm\bf k}(R)_{{\rm\bf k}^\prime, {\rm\bf k}}=\left\{ \begin{array}{ll}1,& \mbox{if~~}{\rm\bf k}^\prime= R{\rm\bf k}-{\rm\bf g}_R \\ 0 & \mbox{otherwise}\end{array}\right.,&D^{\bm\tau}(R)_{{\bm\tau}_\alpha^\prime, {\bm\tau}_\alpha}=\left\{ \begin{array}{ll}1,& \mbox{if~~}{\bm\tau}_\alpha^\prime= \{R|{\rm\bf v}\}{\bm\tau}_\alpha+{\rm\bf t}^\prime \\ 0 & \mbox{otherwise}\end{array}\right..\\
D^{\bm{\tau},\mu}(R)_{{\bm \tau}_\alpha^\prime k,{\bm \tau}_\alpha l}=D^{\bm{\tau}}(R)_{{\bm \tau}_\alpha^\prime {\bm \tau}_\alpha} D^{\mu}(R^\prime)_{kl},\quad& R^\prime=R_{{\bm \tau}_\alpha^\prime}^{-1}RR_{{\bm \tau}_\alpha}\in\tilde{\rm G}^\tau.  \label{eqn:atomic_rep}
\end{array}
\end{align}
Here ${\rm\bf g}_R$ can be zero or a reciprocal lattice vector. The condition of non-zero ${\rm\bf g}_R$ can only occur on the high symmetry points(HSP), lines and planes on the surface of BZ, producing a non unity {\em gauge term} $\exp\{i{\rm\bf g}_R\cdot{\bm \tau}_{\alpha^\prime}\}$. $R_{{\bm \tau}_\alpha}$ is the chosen left coset representative associated with ${\bm \tau}_\alpha$ in the decomposition of the isogonal point group with respect to the $\tilde{\rm G}^\tau$. 

IBRs are defined as BR which are band irreducible for a given connected band topology\cite{Zhang_J:2021}. EBRs occurring in a connected bands are IBR themselves. Whilst the vector spaces spanned by EBR induced from Wyckoff position with multiplicity are reducible for ${\rm\bf k}$ in the interior of BZ, the full band connectivity preclude invariance of any linear combinations on the $\tau_\alpha$ index at HSPs on the surface of BZ. 

In real space, the action of $g\in\mathcal{G}$ permutes the crystallographic orbits of equivalent Wyckoff positions with multiplicity. One expect band invariant subspaces arising from linear combination of EBR components induced from Wannier functions centred on such orbits. With the exception of HSPs on the surface of BZ where the gauge term $\exp\{i{\rm\bf g}_R\cdot{\bf \tau}_{\alpha^\prime}\}$ takes non-unity values, the corresponding linear combination of components of EBR are band invariant sub-spaces. Such invariance are broken at HSPs on the surface of BZ when the band topology is fully connected. However, in the case of split EBRs or CBRs with dynamic interaction, band invariance among such linear combination can be restored if the corresponding bands are connected to nodes of appropriate symmetry (irrep) at these HSPs. These dynamically occurring IBRs are rooted in real space symmetry (permutation of crystallographic orbits of equivalent Wyckoff positions). They are band invariant subspaces of EBRs occurring only as consequence of interaction for a given band topology (i.e. they are not intrinsic properties of the basis).

Based on the transformation properties at these HSPs on the surface of BZ and how the invariance of linear combination of EBR components is recovered there, the IBR can be classified. When the whole EBR appears in a connected band, the EBR itself is a type 1 IBR with the band transformation properties determined by band connectivity at HSPs on the surface of BZ. When the band invariance of linear combination is restored under appropriate band topology (connection to the correct irreps at these HSPs), type 2 IBR is obtained. The transformation properties of the components (combining to form IBRs) of the associated EBR are given by Eq.\eqref{eqn:EBR} with the {\em gauge term removed}. Type 2 IBR is always equivalent to some EBR induced from Wannier functions centred on Wyckoff position with no multiplicity. The type 2 IBR may be further classified as type 2$\alpha$ when it occurs as part of a decomposable EBRs or type 2$\beta$ when interaction with other EBR among CBR is required. Type 3 IBR occurs together with type 2$\alpha$. The vector space spanned by an EBR is the direct sum of type 2$\alpha$ IBRs and type 3 IBR under the split configuration of band connectivity. The transformation properties of type 3 IBRs are determined by Eq.\eqref{eqn:EBR} on the interior of BZ. On the surface of BZ where the gauge term is not unity, the transformation properties of type 3 IBR is determined by the symmetry compliant split band connectivity which also defines the corresponding type 2$\alpha$ IBRs. Type 3 IBRs are generally not equivalent to any other EBRs. The decomposition of some IBRs at HSPs are listed in Appendix C of Ref.\onlinecite{Zhang_J:2021}.


 
Some remarks about the EBR and IBR:
\begin{enumerate}
\item The occurrence of IBR depends on the topology of the associated band, and hence the interaction. They are not a static, intrinsic properties of associated EBRs but dynamic in nature. For example, the EBR is a type 1 IBR only when the associated bands are fully connected. But an EBR may also admit band invariant subspaces (IBRs) under symmetry compliant split band connectivity.
\item Type 2 IBRs are always equivalent\footnote{Equivalence between BR only refers to transformation properties and does not imply linear dependence\cite{Zhang_J:2021}.} to some EBR with Wannier functions centred on Wyckoff position with no multiplicity.
\item Type 2$\beta$ IBR may arise from `indecomposable' EBRs in the context of CBR and interactions. For example, type 2$\beta$ IBR can arise from EBR induced from s orbital centred on orbits of Wyckoff position 2b of the honeycomb lattice ($\mathcal{G}=80({\rm L})$) in a sp$_2$ hybridised system. (see Sec.IV.B of Ref\onlinecite{Zhang_J:2021}).
\item Type 3 IBR are generally not equivalent to any other EBR.
\item The rotational part of the transformation properties of any IBR (including type 1 IBR which is a fully connected EBR) contains no explicit ${\rm\bf k}$ dependence. This enables PIBC with respect to such bases to form a representation of $\mathcal{G}$.
\end{enumerate}

When time-reversal symmetry or magnetic space group is considered, band co-representation \cite{Evarestov_R_A:1989} and corresponding IBRs are used.

\section{Link between the topological invariant of physical dispersion surfaces and IBRs}
\label{sec:ChernEBR}
From the definition of $\phi_{\rm B}$ in Eq.\eqref{eqn:BWZphase}, the transformation properties of its elements such as $\left|\psi_n({\rm\bf k})\right>$ are generally ${\rm\bf k}$ dependent. It is not obvious how such path integral would transform under space group symmetry, or indeed if it forms a representation of the space group. We seek to evaluate $\phi_{\rm B}$ under the TB model in terms of its EBR bases initially. In particular, it is recognised that the transformation properties of EBRs with the Slater Koster phase contains no explicit ${\rm\bf k}$ dependence\cite{Zhang_J:2021} on the rotational part. The TB model provides the link between the EBRs/IBRs and physical electron energy dispersion surfaces of an {\em infinite solid}. This allows the classification of the topology of multi-band energy dispersion surfaces (the vector bundle), described by the TB model and its symmetry compliant Hamiltonian $H({\rm\bf k})$. The approach here make use of full eigenstate $\left|\psi_n({\rm\bf k})\right>$, defined over the full BZ manifold, in contrast to the cell periodic part of the eigenfunction $u_{n, {\rm\bf k}}({\rm\bf r})$ which is only defined in the representation domain. As the un-contractable closed path often reaches all parts of the BZ, the use of full eigenstate and full group method is necessary.

Let $\left|\phi_j({\rm\bf k})\right>$ be the EBR/IBR bases of the TB Hamiltonian with $j$ as a composite index to the BR (There may be multiple EBRs with orbitals of different symmetries of the site symmetry group and located on different Wyckoff positions). The range of $m, n$ are the same as that of $j$. At a given wavevector ${\rm\bf k}$, we have
\begin{equation}
\left|\psi_m({\rm\bf k})\right>=\sum_jU({\rm\bf k})_{jm}\left|\phi_j({\rm\bf k})\right>.
\label{eqn:unitary}
\end{equation}
where $U({\rm\bf k})$ is a ${\rm\bf k}$ dependent unitary matrix and smooth function of ${\rm\bf k}$. Alternatively, we may regard $U({\rm\bf k})^{-1}$ as defining the localised Wannier functions centred on the Wyckoff positions from the energy eigenstates through Fourier Transforms\cite{Marzari_N:2012}. 

We evaluate $\phi_{\rm B}$ by making use of Eq.\eqref{eqn:unitary} for the eigenstates of the Hamiltonian.
\[
\nabla_{\rm\bf k}\left|\psi_n({\rm\bf k})\right>=\sum_j\left[\nabla_{\rm\bf k}U({\rm\bf k})\right]_{jn} \left|\phi_j({\rm\bf k})\right>+\sum_jU({\rm\bf k})_{jn}\nabla_{\rm\bf k}\left|\phi_j({\rm\bf k})\right>
\]
\begin{align}
i\phi_B&=\sum_n\oint_{{\rm\bf k}_0}^{{\rm\bf k}_0}\left<\psi_n({\rm\bf k})\mid\nabla_{\rm\bf k}\mid\psi_n({\rm\bf k})\right>\cdot{\rm d \bf k} \notag \\
&=\sum_n\sum_i\sum_j\oint_{{\rm\bf k}_0}^{{\rm\bf k}_0}U({\rm\bf k})_{in}^\ast\left<\phi_i({\rm\bf k})\right|\left\{\left[\nabla_{\rm\bf k}U({\rm\bf k})_{jn}\right] \left|\phi_j({\rm\bf k})\right>+U(({\rm\bf k})_{jn}\nabla_{\rm\bf k}\left|\phi_j({\rm\bf k})\right>\right\}\cdot{\rm d}{\rm\bf k}\notag \\
&=\sum_n\sum_i\sum_j\oint_{{\rm\bf k}_0}^{{\rm\bf k}_0}\left\{U({\rm\bf k})_{ni}^\dag\nabla_{\rm\bf k}U({\rm\bf k})_{jn}\left<\phi_i({\rm\bf k})\mid\phi_j({\rm\bf k})\right>+U({\rm\bf k})^\dag_{ni}U({\rm\bf k})_{jn}\left<\phi_i({\rm\bf k})\mid\nabla_{\rm\bf k}\mid\phi_j({\rm\bf k})\right>\right\}\cdot{\rm\bf dk}\notag \\
&=\oint_{{\rm\bf k}_0}^{{\rm\bf k}_0}\Tr\left[U({\rm\bf k})^\dag\cdot\nabla_{\rm\bf k}U({\rm\bf k})\right]\cdot{\rm\bf dk}+\sum_j\oint_{{\rm\bf k}_0}^{{\rm\bf k}_0}\left<\phi_j({\rm\bf k})\mid\nabla_{\rm\bf k}\mid\phi_j({\rm\bf k})\right>\cdot{\rm\bf dk} 
\label{eqn:derivation}
\end{align}
where in the last step, we carried out the summation over $i, j$ and making use of orthogonality of the BR (TB) bases in the first term and summation over $n$ and $i$ and making use of unitarity of $U$ for the second term. Consider the first term in the integrand, a vector field in the BZ with its three cartesian components given by,
\[
\Tr \left[U({\rm\bf k})^\dag\frac{\partial U({\rm\bf k})}{\partial {\rm\bf k}_x}\right], \Tr \left[U({\rm\bf k})^\dag\frac{\partial U({\rm\bf k})}{\partial {\rm\bf k}_y}\right],\Tr \left[U({\rm\bf k})^\dag\frac{\partial U({\rm\bf k})}{\partial {\rm\bf k}_z}\right].
\]

Looking at the z component of the curl of this complex vector field
\begin{align*}
Z_3&=\frac{\partial}{\partial {\rm\bf k}_x}\Tr \left[U({\rm\bf k})^\dag\frac{\partial U({\rm\bf k})}{\partial {\rm\bf k}_y}\right]-\frac{\partial}{\partial {\rm\bf k}_y}\Tr \left[U({\rm\bf k})^\dag\frac{\partial U({\rm\bf k})}{\partial {\rm\bf k}_x}\right]\\
&=\Tr \left\{U({\rm\bf k})^\dag\left[\frac{\partial^2U({\rm\bf k})}{\partial{\rm\bf k}_x\partial{\rm\bf k}_y}-\frac{\partial^2U({\rm\bf k})}{\partial{\rm\bf k}_y\partial{\rm\bf k}_x}\right]+\left[\frac{\partial U({\rm\bf k})^\dag}{\partial{\rm\bf k}_x}\frac{\partial U({\rm\bf k})}{\partial{\rm\bf k}_y}-\frac{\partial U({\rm\bf k})^\dag}{\partial{\rm\bf k}_y}\frac{\partial U({\rm\bf k})}{\partial{\rm\bf k}_x}\right]\right\} \\
&=\Tr\left[\frac{\partial U({\rm\bf k})^\dag}{\partial{\rm\bf k}_x}\frac{\partial U({\rm\bf k})}{\partial{\rm\bf k}_y}-\frac{\partial U({\rm\bf k})^\dag}{\partial{\rm\bf k}_y}\frac{\partial U({\rm\bf k})}{\partial{\rm\bf k}_x}\right]
\end{align*}
Since $U({\rm\bf k})$ is unitary,
\[
\frac{\partial U({\rm\bf k})U({\rm\bf k})^\dag}{\partial {\rm\bf k}_x}=\frac{\partial U({\rm\bf k})}{\partial {\rm\bf k}_x}U({\rm\bf k})^\dag+U({\rm\bf k})\frac{\partial U({\rm\bf k})^\dag}{\partial {\rm\bf k}_x}=\mathbb{0}\mbox{~~}\Rightarrow\mbox{~~}\frac{\partial U({\rm\bf k})^\dag}{\partial {\rm\bf k}_x}=-U({\rm\bf k})^\dag\frac{\partial U({\rm\bf k})}{\partial {\rm\bf k}_x}U({\rm\bf k})^\dag
\]
Substitute for $\frac{\partial U({\rm\bf k})^\dag}{\partial {\rm\bf k}_x}$ and $\frac{\partial U({\rm\bf k})^\dag}{\partial {\rm\bf k}_y}$, we obtain
\[
Z_3=-\Tr\left[U({\rm\bf k})^\dag\frac{\partial U({\rm\bf k})}{\partial {\rm\bf k}_x}U({\rm\bf k})^\dag \frac{\partial U({\rm\bf k})}{\partial{\rm\bf k}_y}-U({\rm\bf k})^\dag\frac{\partial U({\rm\bf k})}{\partial {\rm\bf k}_y}U({\rm\bf k})^\dag\frac{\partial U({\rm\bf k})}{\partial{\rm\bf k}_x}\right]=\Tr\left[AB-BA\right]
\]
where $A=U({\rm\bf k})^\dag\frac{\partial U({\rm\bf k})}{\partial{\rm\bf k}_x}$ and $B=U({\rm\bf k})^\dag\frac{\partial U({\rm\bf k})}{\partial{\rm\bf k}_y}$.  Since trace of product of two matrices is insensitive to the order of the matrices, we have
\[
Z_3=Z_1=Z_2=0\mbox{~~~~~}\Rightarrow\mbox{~~~~~}\nabla_{\rm\bf k}\times\Tr \left[\nabla_{\rm\bf k}U({\rm\bf k})^\dag\cdot U({\rm\bf k})\right]=\bm{0}.
\]
Hence  the contribution from this first term in Eq.\eqref{eqn:derivation} is zero by Stokes theorem. Therefore the BWZ phase $\phi_{\rm B}$ is unaffected by the interaction introduced by the Hamiltonian and we have
\begin{equation}
i\phi_B=\mbox{Tr}\underbrace{\oint_{\gamma}\left<\psi_m({\rm\bf k})\mid\nabla_{\rm\bf k}\mid\psi_n({\rm\bf k})\right>\cdot{\rm d \bf k}}_{\mbox{CPIBC for connected bands}}=\mbox{Tr}\underbrace{\oint_{\gamma}\left<\phi_i({\rm\bf k})\mid\nabla_{\rm\bf k}\mid\phi_j({\rm\bf k})\right>\cdot{\rm d \bf k}}_{\mbox{CPIBC for IBR/EBR bases}}.
\label{eqn:realtoEBR}
\end{equation}
It is important to note the implicit assumption that the EBR or IBR bases are complete and invariant under the space group operations.

We further consider a gapped system under the TB model with connected bands (vector bundles) with eigenstates $\left|\psi^A_p({\rm\bf k})\right>$ below and $\left|\psi^B_q({\rm\bf k})\right>$ above the gap. A subset of symmetry compliant Hamiltonian $\hat{\mathcal{H}}({\rm\bf k})\subset \hat{H}({\rm\bf k})$ maintains the gap. Naturally, $\left<\psi^A_p({\rm\bf k})\mid\psi^B_q({\rm\bf k}^\prime)\right>=0,\: \forall {\rm\bf k},{\rm\bf k}^\prime \in \mbox{BZ}$. 

Focusing on one of the connected band, say $\left|\psi^A_p({\rm\bf k})\right>$, we further restrict ourselves to the situation where it posses {\em a complete set of dynamically occurring IBR bases} $\left|\zeta^A_i({\rm\bf k})\right>$ {\em with the membership of the set independent of ${\rm\bf k}$}. The physical intuition behind this restriction is the lack of band inversion in topologically trivial phase. These IBRs can be any of the three types of dynamically occurring IBRs subject to the restriction on the range of hopping parameter which maintain the gap and unitarity of transformation which give rise to them under the full TB model for all bands. Clearly, the space spanned by the original bases $\left|\phi_i({\rm\bf k})\right>$ is divided into invariant subspaces spanned by $\left|\zeta^A_p({\rm\bf k})\right>\oplus\left|\zeta^B_q({\rm\bf k})\right>$. Given the ${\rm\bf k}$ independent set of IBR bases $\left|\zeta^A_i({\rm\bf k})\right>$, $\left|\zeta^B_i({\rm\bf k})\right>$ and their band irreducible nature, there must be a ${\rm\bf k}$ independent similarity transformation that block diagonalise $\hat{\mathcal{H}}({\rm\bf k})$ subject to maintenance of the gap. 

Provided {\em the gap is maintained}, the membership of the IBR bases $\left|\zeta^A_p({\rm\bf k})\right>$ will remain constant independent of the constrained hopping parameters. The same analysis above can be applied to the connected band $\left|\psi^A_p({\rm\bf k})\right>$ yielding Eq.\eqref{eqn:realtoEBR} for relation of $\phi_{\rm B}$ of $\left|\psi^A_p({\rm\bf k})\right>$ and $\phi_{\rm B}^{\rm IBR}$ of its basis $\left|\zeta^A_i({\rm\bf k})\right>$. The requirement of $\left|\phi_i({\rm\bf k})\right>\simeq\left|\zeta^A_p({\rm\bf k})\right>\oplus\left|\zeta^B_q({\rm\bf k})\right>$ implies that the necessary condition of equality of decomposition in terms of irrep of little co-group of ${\rm\bf k}$ between the two set of bases (EBR and IBR). In absence of the assertion of the existence of IBR bases, Eq.\eqref{eqn:realtoEBR} cannot be obtained and selection rule cannot be obtained in the succeeding sections.

Eq.\eqref{eqn:realtoEBR} is clearly the motivation to work with IBRs for the purpose of classification of topologically trivial phase. IBRs are band invariant/irreducible in ${\rm\bf k}$ space and considered the {\em building block for topologically trivial phase} in the sense of the topological invariant as well as IBR bases of the TB Hamilton. This is in contrast to the often quoted statement of EBR as building block of topological quantum chemistry. The summation of indices on the RHS of Eq.\eqref{eqn:realtoEBR} is associated with real space indices of BR (equivalent Wyckoff position, site symmetry group irrep labels). There is no degeneracy concept associated with the IBRs. Thus the potential SU(N) gauge group along the path in physical bands is avoided subject to the restriction on maintenance of gap and existence of a set of IBR bases whose membership is independent of ${\rm\bf k}$. 
 
Given a crystal with space group and EBRs/IBRs, selection rules can be developed for the $\mathcal{BWZ}(\gamma)_{i,j}^{\rm IBR}$ and hence its trace $\phi_{\rm B}^{\rm IBR}$. This can then be applied to physical connected bands using Eq.\eqref{eqn:realtoEBR} if such bands have the a set of IBR bases independent of ${\rm\bf k}$. Some remarks concerning Eq.\eqref{eqn:realtoEBR}.
\begin{enumerate}
\item Derivation of Eq.\eqref{eqn:realtoEBR} only requires the TB model with EBRs/IBRs as bases and independent of the choice of closed loop $\gamma$ in the BZ.
\item For Eq.\eqref{eqn:realtoEBR} to be applicable to a set of connected bands in a gapped system, they must have a set of IBR $\left|\zeta^A_i({\rm\bf k})\right>$ bases with the membership of the set independent of ${\rm\bf k}$ and the gap maintained.
\item If connected bands have the same irrep labels as decomposition of IBR bases at HSPs with the same connectivity and gap remains, then the actual bases of the connected band is complete and closed due to the von Neumann-Wigner (anti-crossing) theorem.
\item If a set of connected bands have a set of IBR bases, then the eigenstates irrep labels at HSPs must be the same as the direct sum of decomposition of the IBR bases. This provides a necessary condition for a set of connected bands to have complete IBR bases (but not sufficient). It is necessary to know the direct sum of decomposition of three different types of IBRs at HSPs as they are the building block of topologically trivial phase.
\item Type 2 IBR is always equivalent to other EBRs arising from Wyckoff position with no multiplicity. Type 1 IBRs are EBRs themselves. Thus if a connected bands have irrep labels as direct sum of decomposition of a set of EBRs and have the same connectivity, then Eq.\eqref{eqn:realtoEBR} hold. This is the basis of symmetry indicator method. However, Type 3 IBR has no equivalent EBRs. Thus symmetry indicator method, based on EBRs, only yield a subset of possible {\em topologically trivial} phase. There is also the additional necessary condition based on reduced tensor element to be checked (see next two sections).
\item The summation of indices on RHS of Eq,\eqref{eqn:realtoEBR} refers to real space (irrep of site symmetry group, Wyckoff position etc) and independent of the presence of degeneracy, band crossing/anti-crossing in ${\rm\bf k}$ space. Hence the degeneracy and SU(N) gauge group problem in the physical connected bands is not present on the RHS.
\item If $\phi_{\rm B}^{\rm IBR}$ from all of the relevant IBRs/EBRs bases in the set are symmetry forbidden, then the $\phi_{\rm B}$ is zero for the physical bands, and the relevant phase of the physical system is symmetry guaranteed to be topologically trivial.
\end{enumerate}
The rest of this manuscript focus on developing symmetry based selection rules for $\mathcal{BWZ}(\gamma)_{i,j}^{\rm IBR}$ and what happens when gap exist in a TB calculation. As with most symmetry derived selection rules, it can only show if a $\phi_{\rm B}$ or Chern number is forbidden by symmetry for a given set of connected bands and restricted range of hopping parameter which maintain the gaps.

\section{Transformation properties of $\mathcal{BWZ}(\lambda)_{i,j}^{\rm EBR}$}
\label{sec:PIBC_tfm}
\begin{figure}[b]
\includegraphics[height=5cm]{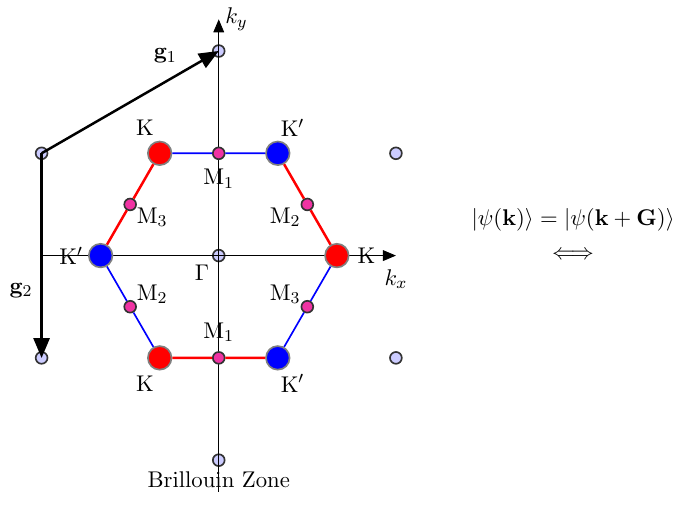}
\includegraphics[height=5cm]{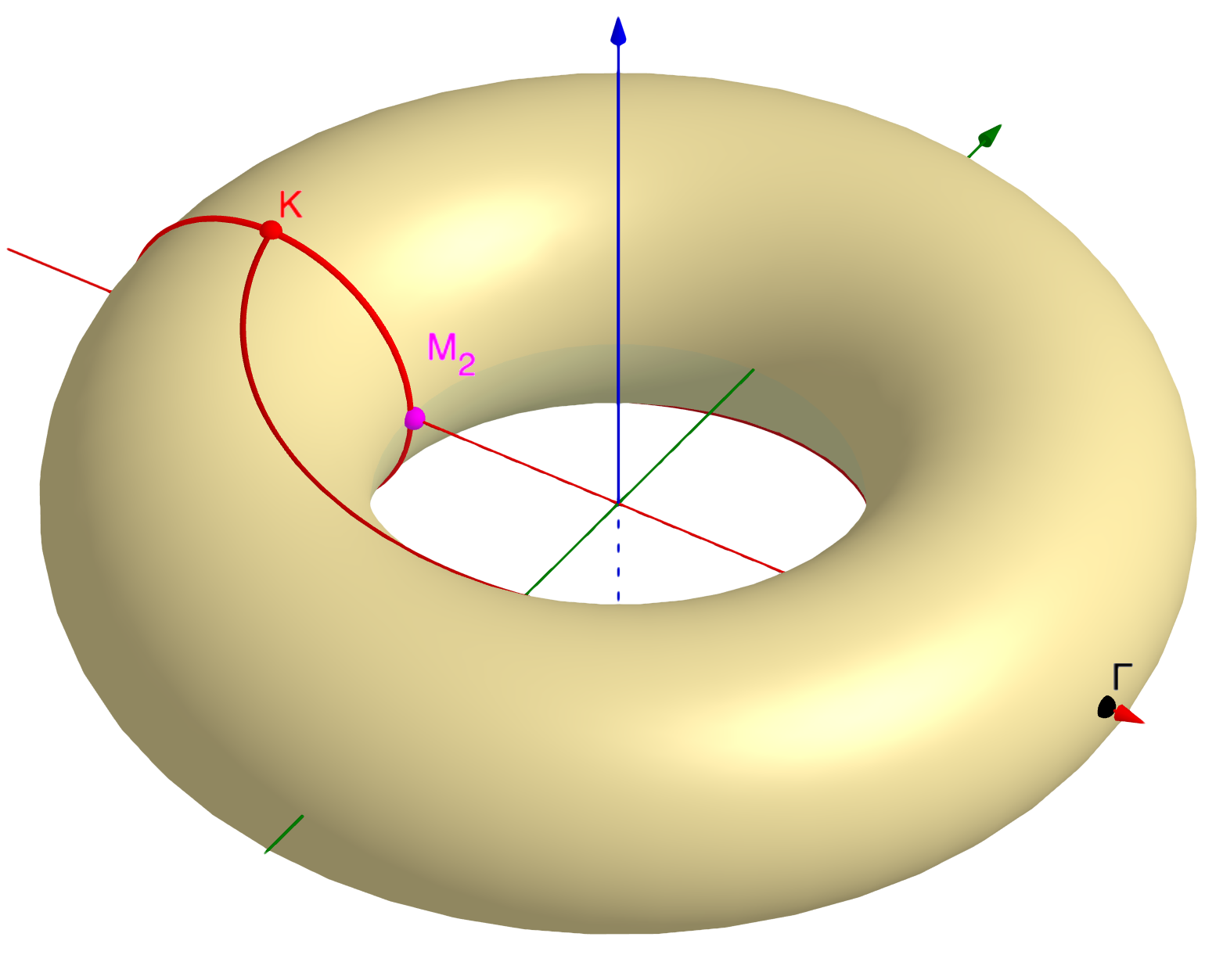}
\caption{Illustration of segments of path integral of BWZ connection with respect to EBRs in the BZ manifold. (a) reduced 1st BZ, (b) BZ torus. The equivalent paths integral $I_{\lambda, mn}$ are taken between $K$ and $K^\prime$ via $M_i, i=1,2,3$ and their inverses. They are illustrated as red and blue lines in (a) and form a representation of the space group $\mathcal{G}$. The paths of the red segments are shown on the torus as in (b). }
\label{fig:BZ}
\end{figure}
It should be noted that when the transformation properties of $\mathcal{BWZ}(\gamma)^{\rm IBR}_{i,j}$ is investigated, the path integral is in the manifold of BZ but the matrix index $\{i,j\}$ are those of Wyckoff position index and irrep labels of the site symmetry group in real space.  Thus one does not have to deal with degeneracy that may arise in connected physical bands. We obtain the general transformation properties of $\mathcal{BWZ}(\lambda)_{i,j}^{\rm IBR}$ and illustrate this with an example in graphene. The location of the paths and HSPs are indicated in Fig.\ref{fig:BZ}.

All elements of $\mathcal{BWZ}(\lambda)_{i,j}^{\rm IBR}$ transform under the action of $\{R|{\rm\bf v+t}\}\in\mathcal{G}$ as per Eq.\eqref{eqn:sym_tfm}, \eqref{eqn:EBR} and its modified form for IBRs. The set of equivalent $\mathcal{BWZ}(\lambda)_{i,j}^{\rm IBR}$ are indexed by $\lambda, i, j$ related to the equivalent paths and composite indices of IBR respectively. Given that $\nabla_{\rm\bf k}$ and ${\rm d{\bf k}}$ transform contragredient to each other and real, $\nabla_{\rm\bf k}\{\bullet\}\cdot{\rm d{\bf k}}=\nabla_{{\rm\bf k}^\prime}\{\bullet\}\cdot{\rm d{\bf k}^\prime}$. $\left|\zeta_n({\rm\bf k})\right>$ `lives' on the path $\lambda$ and belong to one of the arms of $\{\ast{\rm\bf k}\}$. Thus there is a homomorphic map between equivalent paths and arms of $\ast{\rm\bf k}\;\forall\; {\rm\bf k}\in\lambda$. Thus the permutation of the $\lambda$ index encode the transformation of ${\rm\bf k}_a, {\rm\bf k}_b, \nu$, $\nabla_{\rm\bf k}\{\bullet\}\cdot{\rm d{\bf k}}$ and permutation among $\ast{\rm\bf k}$ in Eq.\eqref{eqn:EBR} or that of the IBRs described by the $D^{\rm\bf k}_{{\rm\bf k}^\prime,{\rm\bf k}}(R)$ term. This can then be simply represented by a permutation matrix denoted by $D^C_{\kappa,\lambda}(R)$ where $\kappa,\lambda$ index the equivalent paths. This leaves the remaining transformation of $\left|\zeta_n({\rm\bf k})\right>$ and $\left<\zeta_n({\rm\bf k})\right|$ as a product of representation Eq.\eqref{eqn:EBR} or its equivalent for IBR (excluding the $D^{\rm\bf k}_{{\rm\bf k}^\prime,{\rm\bf k}}(R)$ term) and its conjugate. They deal with the transformation among the $\{i,j\}$, composite indices of the EBR/IBR. It is clear that any translational part depending on $\exp[-iR{\rm\bf k}\cdot({\rm\bf v+t})]$ is removed due to the product of conjugate terms. Thus
\begin{equation}
{\mathcal{BWZ}(\lambda)_{i,j}^{\rm IBR}}=\sum_{\kappa,m,n}D^C_{\kappa, \lambda}(R){D^{\Gamma_\phi}_{mi}}(R)^\ast D^{\Gamma_\phi}_{nj}(R)\mathcal{BWZ}(\kappa)_{m,n}^{IBR}.\label{eqn:PIBC_tfm}
\end{equation}
Here $D^{\Gamma_\phi}_{nj}(R)$ are the transformation properties of IBR of the connected bands. As $D^{\Gamma_\phi}(R)$ does not contain explicit ${\rm\bf k}$ dependance, these $\mathcal{BWZ}(\lambda)_{i,j}$ indeed form representation of the group. The only ${\rm\bf k}$ related terms is connectivity dependent representation matrix where the gauge term $\exp\{i{\rm\bf g}_R\cdot \tau_\alpha^\prime\}$ takes non-unity value on the surface of BZ\cite{Zhang_J:2021}. So one can break $\lambda$ into components $\lambda_\parallel$ which is contained in the surface of BZ and $\lambda_\perp$ which is contained in the interior of the BZ. The HSPs on the surface of the BZ has zero measure and does not contribute to any path integral.

The following takes graphene as an example. For $\gamma$ passing through $K$ and $M_i$ points, there are six equivalent un-contractable closed PIBC for EBR/IBR bases: $K\rightarrow M_r\rightarrow K^\prime\rightarrow M_s\rightarrow K, r,s=1\ldots3, r\ne s$ and three equivalent contractable loops when $r=s$. We shall define these as
\begin{align}
U_{\gamma,mn}&=\int_{\gamma}\left<\phi_m({\rm\bf k})\mid\nabla_{\rm\bf k}\mid\phi_n({\rm\bf k})\right>\cdot{\rm d \bf k}\quad\quad \mbox{\begin{tabular}{c|cccccc}$\gamma$&1&2&3&4&5&6\\\hline $M_{r,s}$ & $M_{2,3}$ & $M_{3,1}$ & $M_{1,2}$ & $M_{3,2}$ & $M_{1,3}$ & $M_{2,1}$  \\ \end{tabular}}\\
Z_{\gamma,mn}&=\int_{\gamma}\left<\phi_m({\rm\bf k})\mid\nabla_{\rm\bf k}\mid\phi_n({\rm\bf k})\right>\cdot{\rm d \bf k}\quad\quad \mbox{\begin{tabular}{c|ccc}$\gamma$&1&2&3\\\hline$M_{r,s}$ & $M_{1,1}$ & $M_{2,2}$ & $M_{3,3}$  \end{tabular}}
\end{align}
These are illustrated in Fig.\ref{fig:BZ}b. (Note the loop $K\rightarrow M_r\rightarrow K^\prime\rightarrow M_s\rightarrow K$ is the same as $K^\prime\rightarrow M_s\rightarrow K\rightarrow M_r\rightarrow K^\prime$ as closed loops with the same sense are one of the same, independent of reference point.) However, it is not difficult to identify that the smaller segment of PIBC for IBR also form a representation of the group. These correspond to the six paths between $K$ and $K^\prime$ via $M_r$ as indicated by the read and blue lines in Fig.\ref{fig:BZ}a. They are defined as
\begin{align}
I_{\lambda,mn}&=\int_{\lambda}\left<\phi_m({\rm\bf k})\mid\nabla_{\rm\bf k}\mid\phi_n({\rm\bf k})\right>\cdot{\rm d \bf k}\quad\quad \mbox{\begin{tabular}{ccccccc}$\lambda$&1&2&3&4&5&6\\\hline${\rm\bf k}_a$ & $K$ & $K$ & $K$ & $K^\prime$ & $K^\prime$ & $K^\prime$ \\ via & $M_1$ & $M_2$ & $M_3$ & $M_1$ & $M_2$ & $M_3$ \\  ${\rm\bf k}_b$ & $K^\prime$ & $K^\prime$ & $K^\prime$  & $K$ & $K$ & $K$ \end{tabular}}
\end{align}
the $\lambda$ index the equivalent paths along which the integration is taken. The six equivalent un-contractable CPIBC and three equivalent contractable loops are simply additions of the relevant segments. For example
\begin{equation}
U_{1,mn}=I_{2,mn}+I_{6,mn},\quad\quad Z_{1,mn}=I_{1,mn}+I_{4,mn}. \label{eqn:EBR_sel}
\end{equation}

In this example, all the paths of $I_{\lambda, i j}$ are contained in the surface of the Brillouin zone. Provided the relevant gauge term involving $\tau_\alpha$ (if present) is included depending on types of IBRs, $D^{\Gamma_\phi}_{nj}(R)$ is well defined and single valued independent of ${\rm\bf k}$. The representation matrix in Eq.\eqref{eqn:PIBC_tfm} is well defined and can be used to construct projection operators in the next section.

\section{Selection rules for $\mathcal{BWZ}(\lambda)^{\rm IBR}_{ij}$ and $\mathcal{BWZ}(\gamma)^{\rm IBR}_{ij}$ }
\label{sec:BWZ_EBR}
This section finds the symmetry permitted form of $\mathcal{BWZ}(\nu)^{\rm IBR}_{ij}$ and develop selection rules on $\phi_{\rm B}^{IBR}$ of $\mathcal{BWZ}(\gamma)^{\rm IBR}_{ij}$ if they exist. The approach follows that of the general matrix element theorem (equivalent of Wigner-Eckart theorem for finite groups)\cite{Koster_G:1958}. A brief description of the methodology can be found in Appendix E of \cite{Zhang_J:2021}. Given that $\mathcal{BWZ}(\lambda)^{\rm IBR}_{ij}$ form a representation of $\mathcal{G}$, the only permitted matrix form of $\mathcal{BWZ}(\lambda)^{\rm IBR}_{ij}$ must transform according to the trivial representation of $\mathcal{G}$. With the representation matrices shown in Eq.\eqref{eqn:PIBC_tfm}, the projection operator for trivial representation is obtained by summing over all elements of $\mathcal{G}$. Because the representation matrices are independent of any translational part, the summation over the translation subgroup is factored out and summation over point group elements (left coset representative to be more precise) is sufficient. 

We take the example of type 1 IBR (fully connected EBR) of p$_z$ orbital centred on the Wyckoff position 2b of graphene ($\mathcal{G}$=80(L)). The p$_z$ orbital transform as $\Gamma_4$ of the site symmetry group D$_{3h}$. Taking the dictionary order of path ($\lambda$), $\tau_\alpha$ of $\left|\phi^{\Gamma_4,2b}\right>$, the projected out form of  $\mathcal{BWZ}(\lambda)^{\Gamma_4,2b}_{ij}$ is given by:
\begin{align}
\arraycolsep=1.6pt
I_{\lambda, ij}=c_1\left(\begin{array}{c|c|c|c|c|c}
\begin{array}{cc} 1 & 0\\0 &1\end{array} & \cdots & \cdots & \cdots & \cdots & \cdots \\ \hline
\cdots & \begin{array}{cc} 1 & 0\\0 &1\end{array} & \cdots & \cdots & \cdots & \cdots \\ \hline
\cdots & \cdots & \begin{array}{cc} 1 & 0\\0 &1\end{array} & \cdots & \cdots & \cdots \\ \hline
\cdots & \cdots & \cdots  & \begin{array}{cc} 1 & 0\\0 &1\end{array} & \cdots & \cdots \\ \hline
\cdots & \cdots & \cdots & \cdots & \begin{array}{cc} 1 & 0\\0 &1\end{array} & \cdots \\ \hline
\cdots & \cdots & \cdots & \cdots & \cdots & \begin{array}{cc} 1 & 0\\0 &1\end{array} 
\end{array}\right)+
c_2\left(\begin{array}{c|c|c|c|c|c}
\begin{array}{cc} 0 & z\\\overline{z} &0\end{array} & \cdots & \cdots & \cdots & \cdots & \cdots \\ \hline
\cdots & \begin{array}{cc} 0 & z\\\overline{z} &0\end{array} & \cdots & \cdots & \cdots & \cdots \\ \hline
\cdots & \cdots & \begin{array}{cc} 0 & z\\\overline{z} &0\end{array} & \cdots & \cdots & \cdots \\ \hline
\cdots & \cdots & \cdots  & \begin{array}{cc}  0 & z\\\overline{z} &0\end{array} & \cdots & \cdots \\ \hline
\cdots & \cdots & \cdots & \cdots & \begin{array}{cc}  0 & z\\\overline{z} &0\end{array} & \cdots \\ \hline
\cdots & \cdots & \cdots & \cdots & \cdots & \begin{array}{cc} 0 & z\\\overline{z} &0\end{array} 
\end{array}\right)
\end{align}
where $c_1, c_2\in \mathbb{C}$ are determined by material parameters and ($z=(1+\sqrt{3}i)/2$). Each of the diagonal blocks corresponds to a equivalent paths indexed by $\lambda$.

Substitute the allowed from of $I_{\lambda,ij}$ into Eq.\eqref{eqn:EBR_sel}, we obtain
\[
Z_{1,mn}=2\left(\begin{array}{cc}c_1 & zc_2\\ \overline{z}c_2 & c_1\end{array}\right)=\mathbb{0}
\]
The anti-Hermitian nature of $Z_{\nu,mn}$ requires $c_1, c_2$ to be purely imaginary. The zero value for contractable loop requires $c_1=c_2=0$. Few more steps lead to
\begin{equation}
U_{\gamma,mn}=\mathbb{0}.
\end{equation}
Taking the trace of these matrices, the BWZ phase $\phi_{\rm B}^{\rm IBR}$ of a {\em completely occupied} type 1 IBR, from Wannier function transforming according $\Gamma_4$ irrep of the site symmetry group and centred on Wyckoff position 2b, is zero and symmetry forbidden. 

The same procedures is repeated for type 1 IBR (fully connected EBR) induced from Wannier functions of all irreps of the site symmetry groups centred on Wyckoff position 1a, 2b, 3d. All results show that $\phi_B^{IBR}$ of these completely occupied type 1 IBRs are symmetry forbidden. In particular, $\phi_B^{IBR}$ of those type 1 IBR arsing from Wyckoff positions 1a (no multiplicity) are symmetry forbidden. Therefore any type 2 IBRs, which are equivalent to type 1 IBR from Wyckoff position with no multiplicity, also have symmetry forbidden $\phi_B^{IBR}$. Since the type 2$\alpha$ IBRs has $\phi_B^{IBR}$ symmetry forbidden, $\phi_B^{IBR}$ of the associated type 3 IBRs are also symmetry forbidden.

It is also possible to look at the transformation properties of $U_{\gamma,ij}$ directly. Similar symmetry permitted form can be obtained and the reduced tensor element is forced to zero because $U_{1,ij}=-U_{4,ij}$. The proof here has chosen a specific integration loop but can be generalised to closed loop of any shape.

These selection rules are not general. i.e. it depends on $\mathcal{G}$ and the IBRs. For example, one cannot derive selection rules for $\mathcal{G}=22$(F222) and orbitals centred on Wyckoff position 4a and 4b. In this case, $\phi_{\rm B}^{\rm IBR}$ is not symmetry forbidden but can still take zero value.

\section{Time reversal symmetry and case for $\mathcal{G}=22, F222$}
\label{sec:G22}
Space group $\mathcal{G}=22, F222$ has been cited as an example where the symmetry does not forbid $\phi_{\rm B}$ of an type 1 IBR (the title of  \cite{Cano_J:2022} refers to ``Topologies invisible to eigenvalues''). In this section, we use the same notation as Ref.\onlinecite{Cano_J:2022} to indicate equivalent paths ($\gamma$ is indicated by the reciprocal lattice vectors). Type 1 IBR (fully connected EBR) with Wannier functions centred on the 4a and 4b Wyckoff positions transforming as trivial representation of the site symmetry group are investigated. These EBRs were shown to be analytically different under periodic boundary conditions, despite the fact there have the same transformation properties\cite{Barcy_H:1988}. If we consider selection rules for $\mathcal{BWZ}(\gamma)^{\rm IBR}_{ij}$ along the closed un-contractable loop $\mathbf{g}_1$ starting at the $\Gamma$ point, and its equivalent generated by the point group $\mathbf{g}_2, \mathbf{g}_3$, and $\mathbf{g}_4=-(\mathbf{g}_1+\mathbf{g}_2+\mathbf{g}_3)$, one can obtain the symmetry permitted form but unable to obtain selection rule based on zero reduced tenor element. Hence the $\phi_{\rm B}^{\rm IBR}$ can take on any value, including zero, as far as symmetry is concerned. This gives a consistent explanation of different $\phi_{\rm B}$ obtainable from distinct but equivalent EBRs with the same decomposition at HSPs\cite{Cano_J:2022}. The equivalence of BR does not necessarily imply equivalence of topological properties. Indeed the work of Barcy et al.\cite{Barcy_H:1988} shows the distinction between the two equivalent EBRs only occur when boundary condition is considered. This is consistent with different $\phi_B$ for these two distinct but equivalent EBRs, the symmetry analysis here and bulk-boundary correspondence.

The situation is different if we bring in time reversal symmetry $T$. The presence of anti-unitary $T$ couples the path $\gamma$ and $-\gamma$ (induced by a transformation in wave-vector space ${\rm\bf k} \mapsto {\rm\bf k}^\prime=-{\rm\bf k}$) as components of the same co-representation of the space group\cite{Bradley_C_J:1968, Evarestov_R_A:1989}. Thus the complete set of CPIBC now includes those identified by $\mathbf{g}_5=-\mathbf{g}_1, \mathbf{g}_6=-\mathbf{g}_2, \mathbf{g}_7=-\mathbf{g}_3$ and $\mathbf{g}_8=-\mathbf{g}_4$. These additional equivalent loops enables selection rules based on zero reduced tensor elements to be constructed. Hence, selection rules can be established for $\phi_B^{IBR}$ of EBR centred on Wyckoff positions 4a and 4b in {\em time-reversal symmetric systems} obeying symmetry of space group $\mathcal{G}=22$.

The same analysis, with CPIBC forming representation of the space group, can be applied to system with magnetic space groups provided the appropriate band co-representation (BCR) \cite{Evarestov_R_A:1989} is used instead of the normal BR. Any symmetry forbidden $\phi_{\rm B}^{IBR}$ (via angular dependent part or reduced tensor elements of CPIBC of BCRs and Eq.\eqref{eqn:realtoEBR}) implies the relevant physical connected bands are topologically trivial.

\section{Further examples of symmetry constraints on $\phi_{\rm B}$ and symmetry protection}
Symmetry compliant tight binding model\cite{Zhang_J:2021} allows proper symmetry analysis of electronic dispersion and application of selection rules if they are present. In particular, the projection operator technique allows establishment of possible connectivity of nodes between HSPs beyond the closed neighbours. In this section, we illustrate the use of symmetry analysis established in the previous two sections in the case of sp$_2$ hybridised bands in graphene (trivial at half or 5/6 occupancy), bands from split EBR of spin-full p$_z$ Wannier function centred on the crystallographic orbits of Wyckoff position 2b in graphene (non-trivial at half occupancy) and bands from split EBR from p$_{xy}$ Wannier functions centred on orbits of Wyckoff position 2b in graphene (trivial at half occupancy). The space group symmetry is that of sub-periodic layer group $\mathcal{G}=80({\rm L})$.

\subsection{Bands from sp$_2$ hybridised orbitals centred on orbits of Wyckoff position 2b}
\label{sec:sp2}
\begin{figure}[t]
\includegraphics[width=0.4\textwidth]{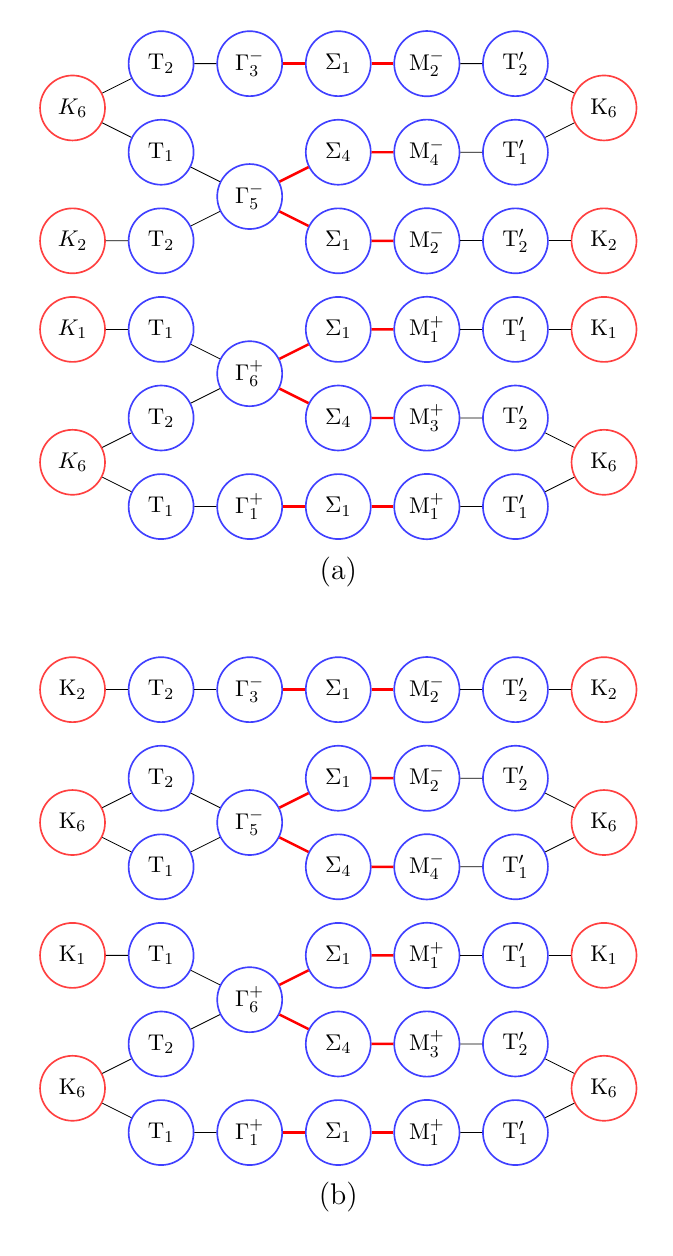} 
\includegraphics[width=0.55\textwidth]{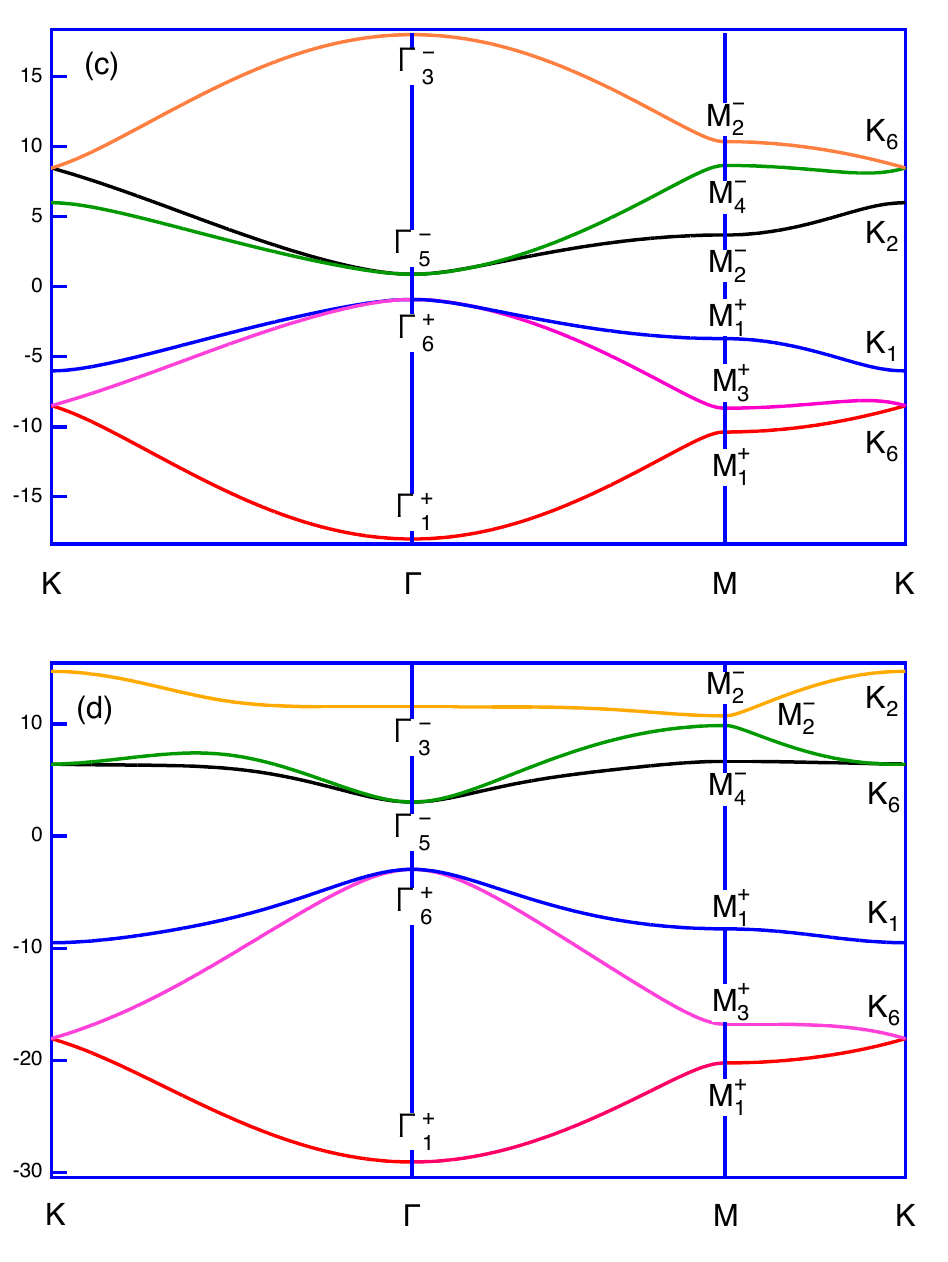}\\
\caption{Symmetry compliant band connectivity of the ${\rm sp_2}$ Wannier functions centred on Wyckoff position 2b. (a) and (b) both exhibit gap at half filling and (b) exhibit additional gap at 5/6 filling. (c) and (d) shows the corresponding dispersion obtained from TB with hooping parameters: (c) $t^{ZNN}_{\Gamma_1,\Gamma_1}=t^{ZNN}_{\Gamma_6,\Gamma_6}=0$, $t^{FNN}_{\Gamma_1,\Gamma_1}=-6$, $t^{FNN}_{\Gamma_1,\Gamma_6}=4$, $t^{FNN}_{\Gamma_6,\Gamma_6:A}=0.3$, $t^{FNN}_{\Gamma_6,\Gamma_6:B}=2$, no second nearest neighbour interaction included. (d) $t^{ZNN}_{\Gamma_1,\Gamma_1}=-8.868$, $t^{ZNN}_{\Gamma_6,\Gamma_6}=0$, $t^{FNN}_{\Gamma_1,\Gamma_1}=-6.769$, $t^{FNN}_{\Gamma_1,\Gamma_6}=5.580$, $t^{FNN}_{\Gamma_6,\Gamma_6:A}=-1.002$, $t^{FNN}_{\Gamma_6,\Gamma_6:B}=4.035$,  $t^{SNN}_{\Gamma_1,\Gamma_1}=0.020$, $t^{SNN}_{\Gamma_1,\Gamma_6:A}=0.050$, $t^{SNN}_{\Gamma_1,\Gamma_6:B}=0.050$, $t^{SNN}_{\Gamma_6,\Gamma_6:A}=0.010$, $t^{SNN}_{\Gamma_6,\Gamma_6:B}=0.210$,  $t^{SNN}_{\Gamma_6,\Gamma_6:C}=0.50i$, $t^{SNN}_{\Gamma_6,\Gamma_6:D}=0.510$. }
\label{fig:sp2}
\end{figure}

The hybridised sp$_2$ orbital forms the backbone of graphene and the associated bands is a known gapped system with trivial topology on either side of the gap under half occupancy. Under the symmetry compliant TB model, the bases are composite EBRs induced from s and p$_{xy}$ orbitals ($\Gamma_1$ and $\Gamma_6$ irreps of the site symmetry group D$_{3h}$) centred on orbits of Wyckoff position 2b. We can order these EBR bases according to the equivalent Wyckoff position index and irrep labels: $\left|{\bm \tau}_A^{2b}, \Gamma_1,1\right>$, $\left|{\bm \tau}_B^{2b}, \Gamma_1,1\right>$, $\left|{\bm \tau}_A^{2b}, \Gamma_6,1\right>$, $\left|{\bm \tau}_A^{2b}, \Gamma_6,2\right>$, $\left|{\bm \tau}_B^{2b}, \Gamma_6,1\right>$,  and $\left|{\bm \tau}_B^{2b}, \Gamma_6,2\right>$. For hybridisation to occur, the difference between the onsite parameters $t^{ZNN}_{\Gamma_1,\Gamma_1}$ and $t^{ZNN}_{\Gamma_6,\Gamma_6}$ is restricted to a certain range depending on the nearest neighbour hopping constants. With these constraints, some of the possible {\em gapped} connectivity configurations are shown in Fig.\ref{fig:sp2} (a) and (b) with the corresponding dispersion relations shown in (c) and (d). The symmetry of eigenstates at HSPs are confirmed by projection operators.

Both configurations of connectivity exhibit gap at 1/2 filling and (b) shows additional gap at 5/6 filling. The irrep labels of nodes at HSPs above and below the gap are shown in Tab.\ref{tbl:sp2_decomposition} together with decomposition of relevant IBRs centred on Wyckoff position 2b. The equivalence in transformation properties between connected bands and identified IBRs centred on Wyckoff position 2b are evident from the same symmetry labels at all HSPs. Thus the Eq.\eqref{eqn:realtoEBR} holds. The cited IBRs are all type 2 and have their equivalence in EBR associated with Wyckoff position 1a. Utilising selection rule established for $\phi_{\rm B}^{\rm IBR}$ of the cited equivalent EBRs in Sec.\ref{sec:BWZ_EBR}, it is clear the $\phi_{\rm B}$ of the connected bands on either side of the gaps are symmetry forbidden and they are symmetry guaranteed to be trivial.

\begin{table}
\caption{Symmetry (irrep labels) at HSPs for the connected bands above and below gap from ${\rm sp_2}$ hybridised orbitals centred on Wykoff position 2b corresponding to band connectivities shown in Fig.\ref{fig:sp2} (a) and (b). The symmetry of relevant IBRs centred on Wyckoff position 2b at these HSPs are listed below.}
\begin{tabular}{|c|c|c|c|} \hline
1/2 filling (a) \& (c)&$\Gamma$ & M & K \\ \hline
Below gap from ${\rm sp_2}$ on 2b&$\Gamma_6^+\oplus\Gamma_1^+$ & $2{\rm M}_1^+\oplus {\rm M}_3^+$ & ${\rm K}_6\oplus {\rm K}_1$ \\ \hline
\begin{tabular}{cc}Type 2 IBRs of $\Gamma_6^+\oplus\Gamma_1^+$ on 2b\\
(Equiv. to EBRs of $\Gamma_6^+\oplus\Gamma_1^+$ on 1a)
\end{tabular}&$\Gamma_6^+\oplus\Gamma_1^+$ & $2{\rm M}_1^+\oplus {\rm M}_3^+$ & ${\rm K}_6\oplus {\rm K}_1$ \\ \hline\hline
Above gap fromf ${\rm sp_2}$ on 2b&$\Gamma_5^-\oplus\Gamma_3^-$ & $2{\rm M}_2^-\oplus {\rm M}_4^-$ & ${\rm K}_6\oplus {\rm K}_2$ \\ \hline
\begin{tabular}{cc}Type 2 IBRs of $\Gamma_5^-\oplus\Gamma_3^-$ on 2b\\
(Equiv. to EBRs $\Gamma_5^-\oplus\Gamma_3^-$ on 1a)\end{tabular} &$\Gamma_5^-\oplus\Gamma_3^-$ & $2{\rm M}_2^-\oplus {\rm M}_4^-$ & ${\rm K}_6\oplus {\rm K}_2$ \\ \hline \hline\hline
5/6 filling (c)&$\Gamma$ & M & K \\ \hline
Below gap from ${\rm sp_2}$ on 2b&$\Gamma_6^+\oplus\Gamma_1^+\oplus\Gamma_5^-$ & $2{\rm M}_1^+\oplus {\rm M}_3^+\oplus{\rm M}_2^-\oplus{\rm M}_4^-$ & $2{\rm K}_6\oplus {\rm K}_1$ \\ \hline
\begin{tabular}{cc}
Type 2 IBRs of $\Gamma_6^+\oplus\Gamma_1^+\oplus\Gamma_5^-$ on 2b\\
(Equiv. to EBR of $\Gamma_6^+\oplus\Gamma_1^+\oplus\Gamma_5^-$ on 1a)\end{tabular}&$\Gamma_6^+\oplus\Gamma_1^+\oplus\Gamma_5^-$ & $2{\rm M}_1^+\oplus {\rm M}_3^+\oplus{\rm M}_2^-\oplus {\rm M}_4^-$ & $2{\rm K}_6\oplus {\rm K}_1$ \\ \hline\hline
Above gap from ${\rm sp_2}$ on 2b&$\Gamma_3^-$ & ${\rm M}_2^-$ & ${\rm K}_2$ \\ \hline
\begin{tabular}{c}Type 2$\beta$ IBRs of $\Gamma_3^-$ on 2b\\(Equiv. to EBR $\Gamma_3^-$ on 1a) \end{tabular}&$\Gamma_3^-$ & ${\rm M}_2^-$ & $ {\rm K}_2$ \\ \hline
\end{tabular}
\label{tbl:sp2_decomposition}
\end{table}
Since the same set of IBRs bases describe the connected bands in the BZ, the $\left|\zeta_i({\rm\bf k})\right>$ bases can be obtained from the eigenstates at $\Gamma$ point using projection operator. They are ordered as $\left| \Gamma_1^+,1\right>$, $\left|\Gamma_6^+,1\right>$, $\left| \Gamma_6^+,2\right>$, $\left|\Gamma_3^-,1\right>$, $\left| \Gamma_5^-,1\right>$,  and $\left| \Gamma_5^-,2\right>$. Their relation to the original EBR  bases associated with the Wyckoff position 2b are given by a ${\rm\bf k}$ independent unitary similarity transform $\mathcal{U}$ (obtained using projection operators at $\Gamma$) as:
\begin{equation}
\mathcal{U}=\frac{1}{\sqrt{2}}\left(\begin{array}{rrrrrr}\phantom{\text{-}}1 & 0 & 0 & 1 & 0 & 0 \\ 1 & 0 & 0 & \text{-}1 & 0 & 0 \\ 0 & 1 & 0 & 0 & 1 & 0 \\ 0 & 0 & 1 & 0 & 0 & 1 \\ 0  & 1  & 0 & 0 &\text{-}1 & 0 \\ 0 & 0 & 1 & 0 & 0 & \text{-}1 \end{array}\right).
\label{eqn:sp2}
\end{equation}
The ${\rm\bf k}$ independence of $\mathcal{U}$ can be easily verified and it block diagonalise the Hamiltonian at all ${\rm\bf k}\in$ BZ {\em subject to maintenance of the gap}. It should be noted that the transformed IBR bases are based on EBRs centred on the Wyckoff positions 2b with multiplicity of 2. But they are equivalent in transformation properties to EBRs centred on Wyckoff position 1a with no multiplicity. i.e. they are dynamically occurring type 2 IBRs subject to the maintenance of the gap.

\subsection{p$_{xy}$ orbital centred on Wyckoff position 2b: Split EBR}
\label{sec:pxy}
\begin{figure}[b]
\includegraphics[width=0.32\textwidth]{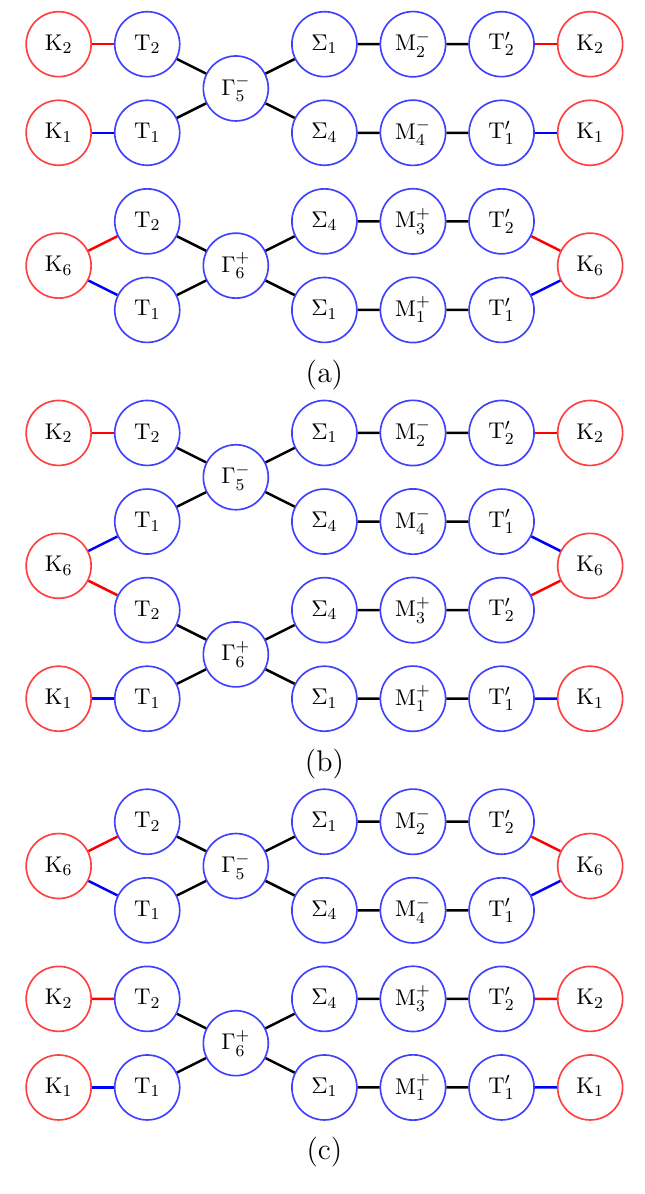}
\includegraphics[width=0.47\textwidth]{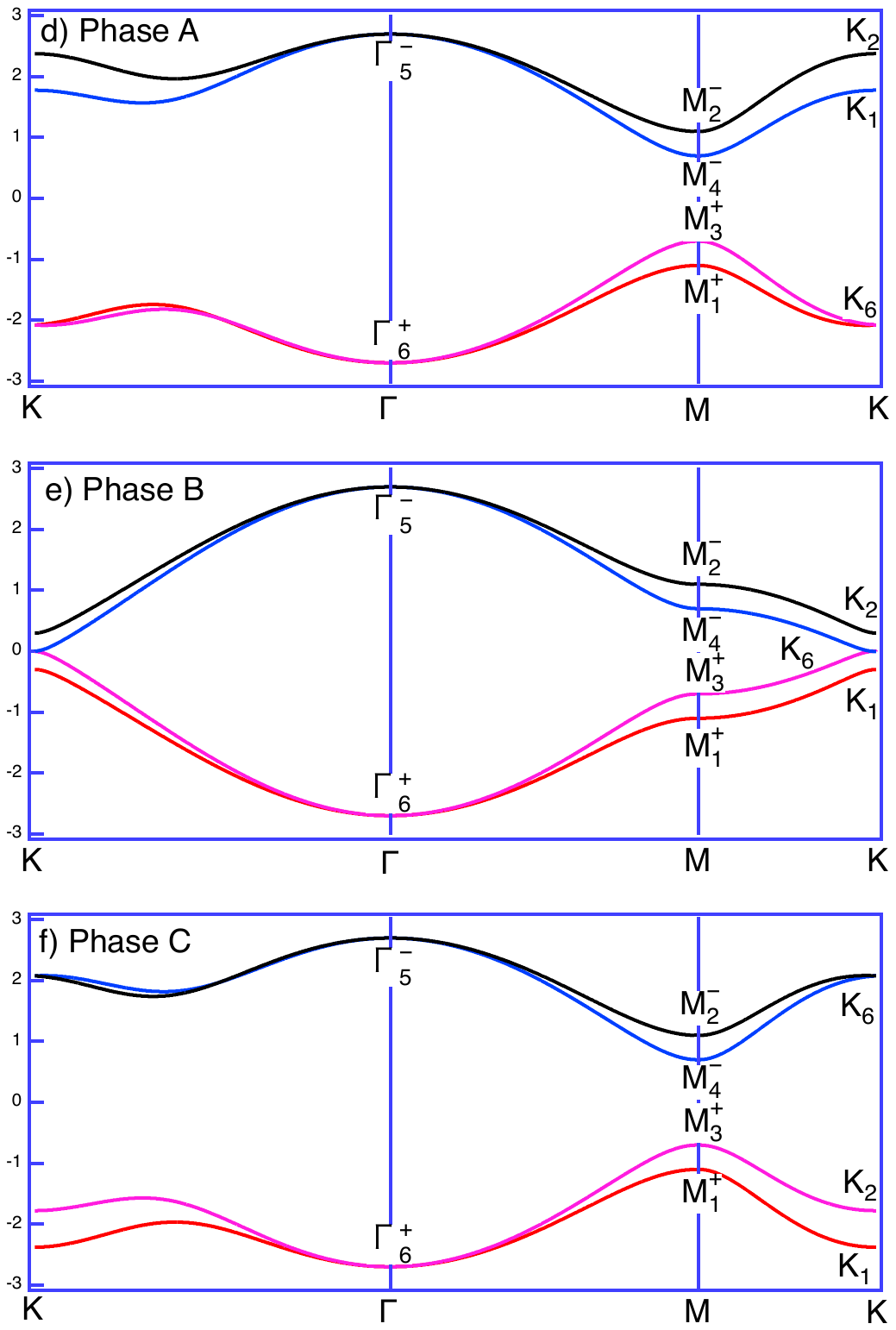}
\caption{Symmetry compliant band connectivities (a) - (c) obtained from projection operator method, and dispersion relations from EBR of $\Gamma_6$ Wannier functions centred on Wyckoff position 2b ($\mathcal{G}=80, p6/mmm(L))$. The dispersion relations are obtained from TB model using hopping parameters: $t^{FNN}_{\Gamma_6,\Gamma_6:A}=0.9$, $t^{FNN}_{\Gamma_6,\Gamma_6:B}=0.1$, $t^{SNN}_{\Gamma_6,\Gamma_6:A}=0$, $t^{SNN}_{\Gamma_6,\Gamma_6:B}=0$, $t^{SNN}_{\Gamma_6,\Gamma_6:C}=0$ for all there phases; and (d) Phase A: $t^{SNN}_{\Gamma_6,\Gamma_6:D}=0.4$, (e) Phase B: $t^{SNN}_{\Gamma_6,\Gamma_6:D}=0$, and (f) Phase C: $t^{SNN}_{\Gamma_6,\Gamma_6:D}=-0.4$.}
\label{fig:pxy}
\end{figure}

The dispersion relation of EBR induced from p$_{xy}$ orbital on Wyckoff position 2b ($\Gamma_6$ of the site symmetry group) is one of the first known to be capable of producing disconnected bands\cite{Cano_J:2018A}. The EBR bases are ordered $\left|{\bm \tau}_A^{2b}, \Gamma_6,1\right>$, $\left|{\bm \tau}_A^{2b}, \Gamma_6,2\right>$, $\left|{\bm \tau}_B^{2b}, \Gamma_6,1\right>$,  and $\left|{\bm \tau}_B^{2b}, \Gamma_6,2\right>$. Fig.\ref{fig:pxy} shows some possible configurations of band connectivities and corresponding dispersion relation obtained from TB calculation using these EBR bases. In particular, phase A \& C are in gapped states whereas phase B is gapless. The connectivity and symmetry labels are identified using projection operator techniques. These are listed in Tab.\ref{tbl:pxy_decomposition} together with decomposition of relevant IBRs centred on orbits of Wyckoff position 2b. 
\begin{table}[b]
\caption{Irrep labels of nodes at HSPs for the bands described by EBR induced from ${\rm p_{xy}}$ Wannier functions localised  on Wyckoff position 2b of lattice with $\mathcal{G}=80({\rm L})$. There are three phases, two of which are gapped.}
\begin{tabular}{|c|c|c|c|} \hline
&$\Gamma$ & M & K \\ \hline
Phase A above gap&$\Gamma_5^-$ & ${\rm M}_2^-\oplus {\rm M}_4^-$ & ${\rm K}_1\oplus {\rm K}_2$\\ \hline
Type 3 IBR of $\Gamma_5^-$ on 2b&$\Gamma_5^-$ & ${\rm M}_2^-\oplus {\rm M}_4^-$ & ${\rm K}_1\oplus {\rm K}_2$\\ \hline
Phase A below gap&$\Gamma_6^+$ & ${\rm M}_1^+\oplus {\rm M}_3^+$ & ${\rm K}_6$ \\ \hline
\begin{tabular}{cc}Type 2$\alpha$ IBR of $\Gamma_6^+$ on 2b\\(Equiv. to $\Gamma_6^+$ EBR on 1a)\end{tabular}&$\Gamma_6^+$ & ${\rm M}_1^+\oplus {\rm M}_3^+$ & ${\rm K}_6$ \\ \hline\hline

Phase B  & $\Gamma_6^+\oplus\Gamma_5^-$ & ${\rm M}_1^+\oplus{\rm M}_3^+\oplus {\rm M}_2^-\oplus {\rm M}_4^-$ & ${\rm K}_6\oplus{\rm K}_1\oplus{\rm K}_2$ \\ \hline
Type 1 IBR of $\Gamma_6$  on 2b & $\Gamma_6^+\oplus\Gamma_5^-$ & ${\rm M}_1^+\oplus{\rm M}_3^+\oplus {\rm M}_2^-\oplus {\rm M}_4^-$ & ${\rm K}_6\oplus{\rm K}_1\oplus{\rm K}_2$ \\ \hline
\hline
Phase C above gap&$\Gamma_5^-$ & ${\rm M}_2^-\oplus {\rm M}_4^-$ & ${\rm K}_6$\\ \hline
\begin{tabular}{cc}Type 2$\alpha$ IBR of $\Gamma_5^-$ on 2b\\(Equiv. to $\Gamma_5^-$ EBR on 1a)\end{tabular}&$\Gamma_5^-$ & ${\rm M}_2^-\oplus {\rm M}_4^-$ & ${\rm K}_6$\\ \hline
Phase C below gap&$\Gamma_6^+$ & ${\rm M}_1^+\oplus {\rm M}_3^+$ & ${\rm K}_1\oplus {\rm K}_2$ \\ \hline
Type 3 IBR of $\Gamma_6^+$ on 2b &$\Gamma_6^+$ & ${\rm M}_1^+\oplus {\rm M}_3^+$ & ${\rm K}_1\oplus {\rm K}_2$ \\ \hline\hline
\end{tabular}
\label{tbl:pxy_decomposition}
\end{table}

The EBR bases associated with p$_{xy}$ orbitals centred on orbits of Wyckoff position 2b is decomposable\cite{Elcoro_L:2017}. The manifestation of this EBR as bands is dependent on the interaction parameters with three distinct atomic limits. The disconnected configuration of phase A and C are associated with IBRs of type 2$\alpha$ and type 3. Phase B correspond to type 1 IBR. The equivalence of these IBR with the symmetry labels of the bands are evident. Given all connected bands are described with IBRs, Eq.\eqref{eqn:realtoEBR} holds and the $\phi_B$ can be evaluated in terms of $\phi_B^{IBR}$. Again the $\phi_B^{IBR}$ of all the relevant IBRs are symmetry forbidden by the reduced tensor element. Therefore $\phi_B$ is symmetry forbidden for all the connected bands in all three phases in Fig.\ref{fig:pxy}.

The transformation properties of type 3 IBR in the interior of BZ is described its EBR components given by Eq.\eqref{eqn:EBR}. On the surface of BZ where the gauge term exist, its transformation properties are determined by the irrep labels of the nodes of EBR after removal of the corresponding type 2$\alpha$ IBR. The same process as Sec.\ref{sec:BWZ_EBR} can be performed for the type 3 IBR here. The associated $\phi_B^{IBR}$ is also symmetry forbidden by the zero reduced tensor element. Despite having no equivalent EBR, the associated $\phi_B$ of the connected band is symmetry forbidden and the bands are guaranteed to be trivial. Such conclusion clearly contradict the basic hypothesis of symmetry indicator method built on EBRs.

The relations of these IBR in phase A and C to the original EBR bases can be obtained from projection operators at $\Gamma$ points. These IBRs are labelled by $\left|\Gamma_6^+,1\right>$, $\left|\Gamma_6^+,2\right>$, $\left|\Gamma_5^-,1\right>$, and $\left|\Gamma_5^-,2\right>$. The unitary similarity transform is given by
\[
\mathcal{U}=\frac{1}{\sqrt{2}}\left(\begin{array}{rrrr}\phantom{\text{-}}1 & \phantom{\text{-}}0 & \phantom{\text{-}}1 & \phantom{\text{-}}0 \\ 0 & 1 & 0 & 1 \\ 1 & 0 & \text{-}1 & 0  \\ 0 & 1 & 0 & \text{-}1  \end{array}\right).
\]
These quantum phases are also symmetry protected as SPTS. Phase A and C are dependent on the sign of second nearest neighbour hopping parameter $t^{SNN}_{\Gamma_6,\Gamma_6:D}$. A transition between the two phases cannot occur without closing the gap as in Phase B.

\subsection{Bands from spin-full p$_z$ orbital centred on orbits of Wyckoff position 2b} 
\begin{figure}
\includegraphics[width=0.51\textwidth]{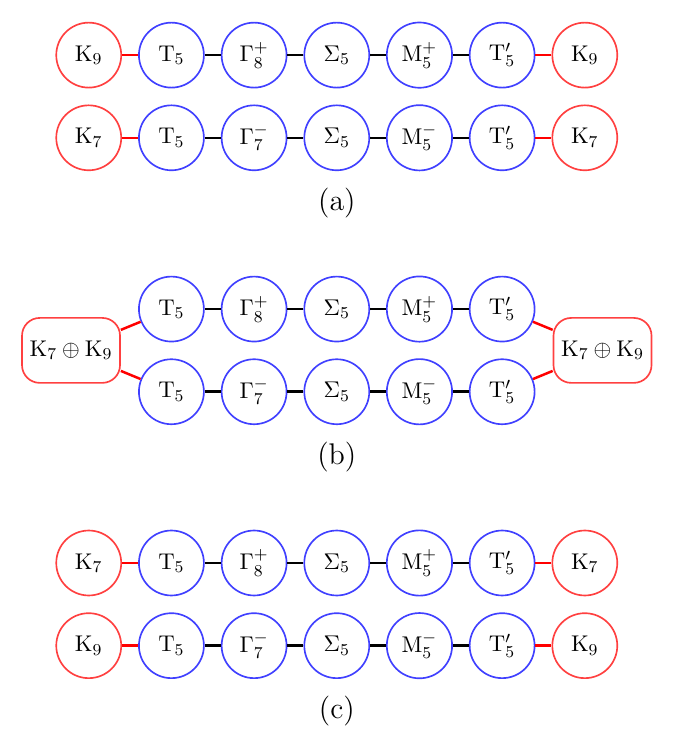}
\includegraphics[width=0.34\textwidth]{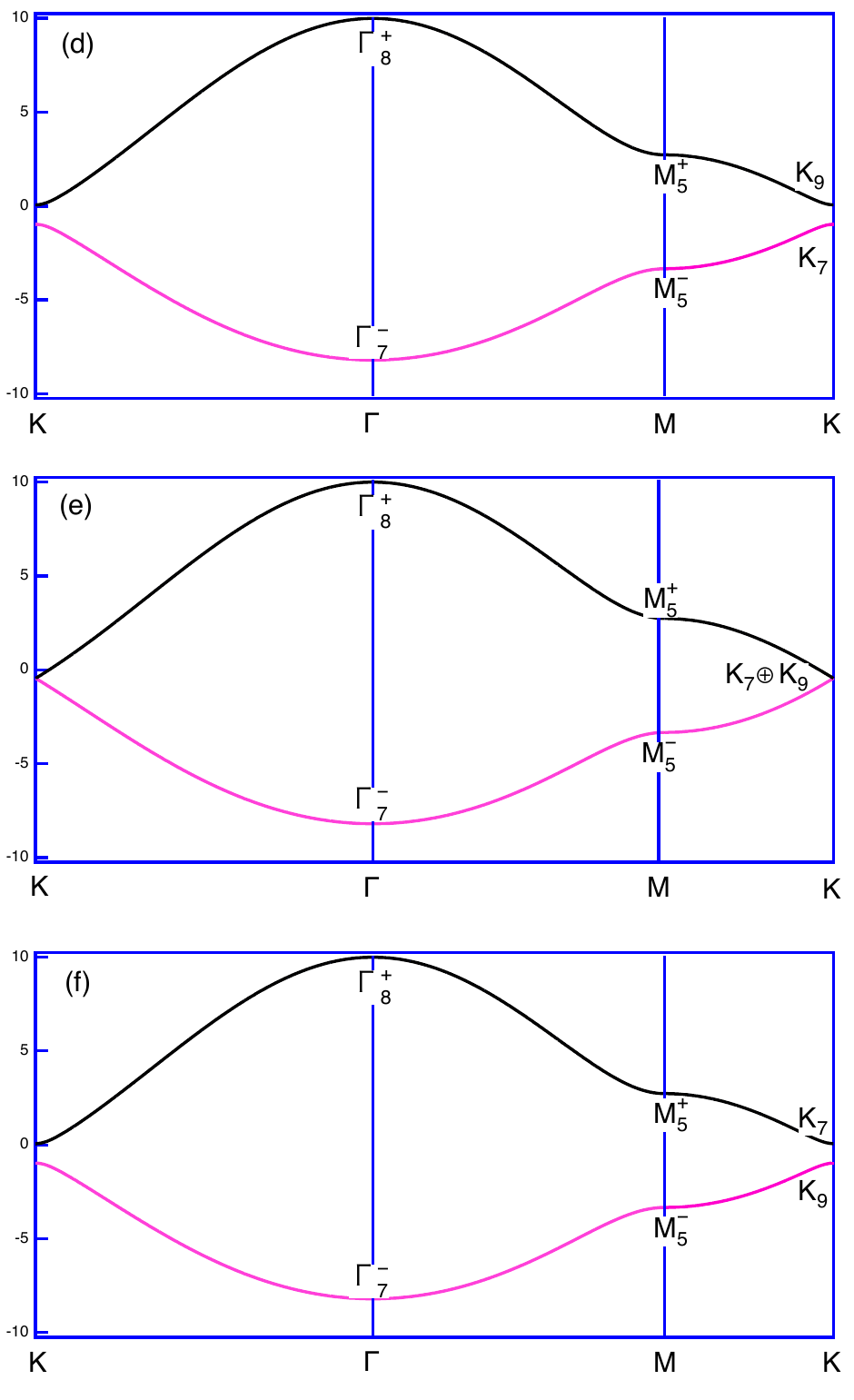}\
\caption{Band connectivities of EBR induced from the spin-full p$_z$ orbital centred on Wyckoff position 2b. (a)-(c) are band connectivities and (d)-(e) are the corresponding dispersion from TB model with the following hopping parameters. $t_{\Gamma_8, \Gamma_8}^{ZNN}=0$, $t_{\Gamma_8, \Gamma_8}^{FNN}=-3.033$, $t_{\Gamma_8, \Gamma_8: A}^{SNN}=0.15$. (d) $t_{\Gamma_8, \Gamma_8:B}^{SNN}=0.1i$, (e) $t_{\Gamma_8, \Gamma_8:B}^{SNN}=0$, and (f) $t_{\Gamma_8, \Gamma_8:B}^{SNN}=-0.1i$.}
\label{fig:pz}
\end{figure}

The last example concerns the spin-full ${\rm p_z}$ orbitals centred on the Wyckoff position 2b. The four EBR bases of the TB Hamiltonian of the system are:  $\left|{\bm \tau}^{2b}_A, \Gamma_8,1\right>$, $\left|{\bm \tau}^{2b}_A, \Gamma_8,2\right>$, $\left|{\bm \tau}^{2b}_B, \Gamma_8,1\right>$,  and $\left|{\bm \tau}^{2b}_B, \Gamma_8,2\right>$. The projection operator technique establish three possible connectivities (two gapped phases and one connected with zero gap) and they are shown in Fig.\ref{fig:pz} together with the corresponding dispersion relation from TB calculation. The symmetry labels of eigenstates at HSPs in the phases D, E, and F are listed in Tab.\ref{tbl:spin_pz}. Connected component bands of phase D \& F are not consistent with any IBR. One may follow the previous examples and try to establish a possible similarity transform $\mathcal{U}$ by diagonalising $H({\bm 0})$. The $\mathcal{U}$ established for both phase D \& F are the same. 
\begin{table}
\begin{tabular}{ccccc} \hline
Phase & Band & $\Gamma$ & M & K \\ \hline
\multirow{2}{*}{D} & CB & $\Gamma_8^+$ & M$_5^+$ & K$_9$ \\ \cline{2-5}
& VB & $\Gamma_7^-$ & M$_5^-$ & K$_7$ \\ \hline
E & - & $\Gamma_7^-\oplus\Gamma_8^+$ & ${\rm M}_5^-\oplus{\rm M}_5^+$ & \underline{${\rm K}_7\oplus{\rm K}_9$} \\ \hline
\multirow{2}{*}{F} & CB & $\Gamma_8^+$ & M$_5^+$ & K$_7$ \\ \cline{2-5}
& VB & $\Gamma_7^-$ & M$_5^-$ & K$_9$ \\ \hline
\end{tabular}
\caption{Irrep labels of eigenstates at HSPs from bands produced by EBR induced from spin-full p$_z$ Wannier function centred on Wyckoff position 2b. No IBR can be identified from disconnected components.}
\label{tbl:spin_pz}
\end{table}

Indeed, such similarity transform can block diagonalise $H({\rm\bf k})$ for most regions of the BZ except in the vicinity of K. Using the projection operator technique, it is not difficult to identify the eigenstates at $\Gamma$ and K in terms of the original EBR bases: 
\[
\left\{
\begin{array}{l}
\left\{\left|\bm{\tau}_A^{2b},\Gamma_8,1\right>+\left|\bm{\tau}_B^{2b},\Gamma_8,1\right>\right\}\mapsto\left|\Gamma_8^+,1\right>\\
\left\{\left|\bm{\tau}_A^{2b},\Gamma_8,2\right>+\left|\bm{\tau}_B^{2b},\Gamma_8,2\right>\right\}\mapsto\left|\Gamma_8^+,2\right>\\
\left\{\left|\bm{\tau}_A^{2b},\Gamma_8,1\right>-\left|\bm{\tau}_B^{2b},\Gamma_8,1\right>\right\}\mapsto\left|\Gamma_7^-,1\right>\\
\left\{\left|\bm{\tau}_A^{2b},\Gamma_8,2\right>-\left|\bm{\tau}_B^{2b},\Gamma_8,2\right>\right\}\mapsto\left|\Gamma_7^-,2\right>
\end{array}\right.
\mbox{~~~and~~~}
\left\{
\begin{array}{l}
\left|\bm{\tau}_A^{2b},\Gamma_8, 2\right>\mapsto\left|{\rm K}_7, 1\right>\\
\left|\bm{\tau}_B^{2b},\Gamma_8, 1\right>\mapsto\left|{\rm K}_7, 2\right>\\
\left|\bm{\tau}_A^{2b},\Gamma_8, 1\right>\mapsto\left|{\rm K}_9, 1\right>\\
\left|\bm{\tau}_B^{2b},\Gamma_8, 2\right>\mapsto\left|{\rm K}_9, 2\right>
\end{array}\right.
\]
Clearly, $\mathcal{U}$ at $\Gamma$ and K cannot be the same and the eigenstates at K always contain IBR components associated with the band on the other side of the gap at $\Gamma$ point. This is entirely inline with early concept of `band inversion' occurring associated with the existence of non-trivial topological phases (eg. $\alpha$-Sn). The `band inversion' here may be taken as incursion of IBR bases across the gap and may occur anywhere in the BZ. Thus ${\rm\bf k}$ independent $\mathcal{U}$ cannot exist to block diagonalise $H({\rm\bf k})$ throughout the BZ. Symmetry cannot be used to forbid the $\phi_{\rm B}$ of the physical dispersion from taking non-zero value in a connected component bands in phase D \& F. They are potentially topologically non-trivial. Again the two phases are separated by phase E which has zero gap due to accidental degeneracy between K$_7$ and K$_9$. 

To produce the transition between the distinct non-trivial phases D \& F, it is necessary to change the sign of hopping parameter $t_{\Gamma_8, \Gamma_8:B}^{SNN}$. Thus such phase transition is not possible without closing the gap (phase E) when $t_{\Gamma_8, \Gamma_8:B}^{SNN}=0$. Thus these topological phases are symmetry protected\cite{Wen_X_G:2004}.

\section{Discussion}
The symmetry constraint on a scaler topological invariant, $\phi_{\rm B}$, can only exist in the form of selection rule. It is either forbidden (trivial for associated vector bundle) or potentially non-zero (potentially non-trivial though does not rule out triviality). The paradigm of the symmetry based searches here is really for topologically trivial phase with other connected bands/Bloch bundles potentially having non-zero BWZ phase. For $\phi_{\rm B}$ to be forbidden, as discussed in Sec.\ref{sec:PIBC_tfm} and \ref{sec:BWZ_EBR}, there are two necessary conditions.
\begin{enumerate}
\item Connected bands having a complete closed set of dynamically occurring IBR bases $\left|\zeta^A_i({\rm\bf k})\right>$ with the membership of the set independent of ${\rm\bf k}$.
\item Selection rule based on zero $\phi_{\rm B}^{\rm IBR}$ for contractable loop and forbidden reduced tensor element (dependent on specific space group) of PIBC for the relevant EBRs/IBRs.
\end{enumerate}
These are the basis for identifying connected bands/Bloch bundles as topologically trivial. Connected bands/Bloch bundles that fails any of the two conditions are potentially non-trivial though the BWZ phase can still take the value zero as far as symmetry is concerned. The following discussion addresses some issues raised in this and the accompanying manuscript\cite{Zhang_J:2021}.
\subsection{Multiple atomic limits for EBR induced from Wyckoff position with multiplicity}
It was understood that EBR induced from Wannier functions centred on Wyckoff position with multiplicity can be decomposable\cite{Elcoro_L:2017}. In this sense, EBR is already not a fundamental building block of connected bands. The original physical intuition from atomic insulators as prototype trivial phase clearly need revision given the existence of multiple atomic limits in  EBRs\cite{Zhang_J:2021} arising from Wyckoff position with mutliplicity. Whilst one may think these multiple atomic limits are the intrinsic properties of the EBRs, the manifestation of any one atomic limits over others are the results of dynamics, i.e. the combination of hopping parameters within the TB framework.  
\subsection{EBR versus IBR}
The dynamically occurring type 2 \& 3 IBR arising from linear combinations of parental EBR components clearly spans smaller band invariant vector space compared to its parental EBR. From the point of view of topologically trivial matter, the IBRs are clearly the building block as full description of connected band by such bases allows the evaluation of BWZ phase in terms of them. As shown in the example in Sec.\ref{sec:sp2}, IBRs can exist on opposite side of the gap whilst its parental EBR appears on both sides. EBR induced from Wannier functions centred on Wyckoff position with multiplicity can have different manifestation of band connectivity depending on the dynamics and therefore would not be suitable building block for topologically trivial phase. The fact that type 2 IBRs are always equivalent to other EBRs centred on Wyckoff position with no multiplicity concealed the existence of IBR basis and the role played by them. 

Not all symmetry permitted connected bands/Bloch bundle can be described by a complete set of IBRs. This naturally leads to topologically non-trivial phase and associated concept of band inversion. Within the TB framework, the bases are EBRs when all the bands are considered together, across the gap if existed. However, it is not necessary for an isolated connected band in a gapped system to be described by an EBR (for example, the band described by type 3 IBR in Fig.\ref{fig:pxy}). It is not clear why EBR takes such a fundamental role in classification of non-trivial phases\cite{Bradlyn_B:2017, Elcoro_L:2020}.

The concept of band inversion, leading to topologically non-trivial phase, is also made clear by the analysis here. The absence of IBR description breaks the validity of Eq.\eqref{eqn:realtoEBR}. One can clearly draw an equivalence of `band inversion' and absence of IBR description of the bands. In other words, incursion of IBRs across the gap means `band inversion'. From the perspective of vector bundles on either side of the energy gap, topologically trivial phases are separate bundles on the same manifold. In contrast, non trivial phase are linked to one another due to absence of IBR description.  

The concept of `fragile topology' has been associated with band whose decomposition at HSPs correspond to that of differences (linear combination with both positive and negative integer coefficients) between EBRs\cite{Elcoro_L:2020}. A particular case need to be considered here is that of type 3 IBR whose direct sum with associated type 2$\alpha$ IBR results in the parental EBR. This would yield the normal fragile topology as it correspond to differences in decomposition of parental EBR and type 2$\alpha$ IBR (which is equivalent to EBR induced from Wyckoff position with no multiplicity). As it is a type 3 IBR, the first necessary condition for selection rule on $\phi_B$ is satisfied. Provided the second condition on reduced tensor element is satisfied, the band described by it is symmetry guaranteed to be topologically trivial. Indeed, it is not clear how to define what is `fragile topology' within the framework of symmetry analysis here.

\subsection{Symmetry indicator method}
Symmetry indicator method appears to be very successful in classification of topological phases with analysis based on EBRs. Yet the work here is based on the paradigm of dynamically occurring IBRs and identification of topologically trivial phase. 

The dynamically occurring type 2 IBRs are always equivalent to EBRs induced from other Wannier functions centred on Wyckoff position with no multiplicity. In a set of gapped topologically trivial bands, the components of parental EBR of a dynamically occurring IBR may appear on both side of the gap. On the issue of distinction between trivial and non-trivial phases based on decomposition at HSPs, the assessment of symmetry indicator method is often the same as analysis here because of the equivalence of type 2 IBR and other EBRs. However, type 3 IBRs, described in phase A and C of Fig.\ref{fig:pxy} with bands connected to K$_1$ and K$_2$ nodes, have no equivalence in other EBRs. Therefore such phase would be topologically non-trivial based on symmetry indicator method. The symmetry analysis based on Wigner-Eckart theorem here suggests otherwise. Another case of failure of symmetry indicator method is that described in Sec.\ref{sec:G22}. In absence of time-reversal symmetry, selection rules on $\phi_B$ cannot be developed due to absence of restriction on the reduced tensor element. Yet the symmetry indicator method would deem any band produced by EBR with Wannier functions centred on Wyckoff position 4a or 4b as topologically trivial. Such prediction of the symmetry indicator method was already contradicted by work of Cano et al.\cite{Cano_J:2022}. In contrast the symmetry analysis in Sec.\ref{sec:G22} gave a consistent explanation of the finding of Ref.\cite{Cano_J:2022}. The work of Barcy et al.\cite{Barcy_H:1988} also shows these EBRs are distinct analytically when boundary condition is imposed. Hence the hypothesis of symmetry indicator method based on EBR with zero $\phi_B$ generally is false.

With the notion of dynamically occurring IBRs, the symmetry analysis performed here suggests a shift of paradigm to search of topologically trivial phase based on IBR descriptions (as building block) and zero reduced tensor element selection rules.

\subsection{Role of HSPs on the surface of BZ}
In the context of necessary conditions for topologically trivial phase, the symmetry of eigenstates at HSPs is absolutely crucial in determining the existence of complete description by IBRs. However, they play a lesser role in determining the reduced tensor element of PIBC. Whilst the gauge term may affect how the components of type 1 IBR transform on high symmetry planes/lines as well as HSPs on the surface of BZ, one can always choose $\gamma$ that do not contain any segment on the surface (i.e. normal to the surface of BZ). Thus the contribution to PIBC from these HSPs have zero integration measure along the path. If such a path is chosen, the transformation properties of contributing integral measure from Eq.\eqref{eqn:PIBC_tfm}{\em in the interior of BZ} is entirely independent of the gauge term. From the perspective of second necessary condition, the reduced tensor element is forbidden only if opposing paths are among equivalent CPIBC under action of $g\in\mathcal{G}$. However, one must recall the essential role the irrep labels on HSPs on the surface of BZ plays in judgement if the first necessary condition is satisfied.

\subsection{Context of K-theory}
In treatment of band connectivity using combinatroics\cite{Kruthoff_J:2017}, connected bands are considered as elements of a ring for which the operation of addition is defined. The elements of the ring also from an abelian group with respect to the addition operation. The element of the ring is identified by the multiplicity of all irreps of group of ${\rm\bf k}$ at HSPs in the RD with connectivity compliant to compatibility relations and its generalised Berry phase. 

The symmetry analysis here is based on the TB framework and existence of IBR bases for the topological trivial phases. It has the following implications under K-theory. First, let's restrict the scope to TB framework in which all the bases of the Hamiltonian are EBRs (having real space representation for all bands) and closed path and its inverse in BZ are both contained in the representation/co-representation of space group formed by CPIBC. In the context of algebraic structure of the ring, the addition operation can be understood in terms of addition of bases of complete EBRs with appropriate division of irreps at HSPs in gapped system. In gapped systems, the Berry phase or topology of the connected component bands depends on the division of irrep multiplicity at HSPs by the gap and band connectivity. The addition of bases for all bands modulo EBR preserve real space description of all bands and avoid the problem of negative multiplicity numbers in gapped system with weak topology\cite{Elcoro_L:2020}.

For topologically trivial gapped system (symmetry forbidden $\phi_B$ for connected component bands), the division of bases across the gap must be along the IBRs, thus preserving the real space description of the connected component bands and selection rule on $\phi_B$. Bands on one side of the gap are the inverse of bands on the other side in the context of the abelian group. They both have zero Berry phase. The identity element is then any bands described by full EBRs. In terms of irrep multiplicity numbers at HSPs, the addition for all bands is modulo EBRs. As an example, bands described by type 3 IBR and accompanying type 2$\alpha$ IBR in gapped system are the representation of inverse elements to each other. The addition of IBR to one side of the gap must be mirrored on the other side to maintain the inverse relation and real space representation of all bands by EBRs. This lead to the conclusion that IBRs are the building block of topological trivial phase.

In gapped non-trivial system under TB regime (no complete IBR bases for connected component bands), bands on either side of the gap are still the inverse of each other. They have opposite Berry phases to each other for a given path in the BZ. The addition operation corresponds to additions of real space bases (EBRs) for all bands with the gap maintained at appropriate places. Beyond compatibility relation, there is no further restriction on the division of bases across the gap as the dimension of BR bases describing a connected component bands exceeds the number of bands. In some sense, the connected bands are a representation of an abelian group with multiple elements in the kernel of the homomorphism. The abelian group here is the addition of Berry phases for a give closed path in the BZ.

It should be noted that a set of multiplicity numbers at HSPs in RD compliant to compatibility relations do not necessarily translate to unique Berry phase. This can be seen in Fig.2 of the accompanying manuscript\cite{Zhang_J:2021}. Both configurations of connectivity are compliant to compatibility relations with the same multiplicity numbers. However, only (a) corresponds to the trivial phase whereas (b) is symmetry indicated as non-trivial. They are distinct elements of the ring with different Berry phase despite having the same multiplicity numbers at HSPs. In the same way, having the same multiplicity numbers at all HSPs is only a necessary condition for equivalence of BRs but not a sufficient one. Qualification of the same band connectivity is required. 

If the reduced tensor element of CPIBC is not forbidden by the space group, then distinct real space representations of $\mathcal{G}$ with the same multiplicity numbers and connectivity may have different Berry phases (see Sec.\ref{sec:G22}). The picture based on addition of bases modulo EBR for all bands no longer work.

\section{Acknowledgement}
The author wishes to acknowledge Department of Physics and Institute for Advanced Studies at Tsinghua University for their generous support of his sabbatical in 2018/19 during which all this work started. Mr. Syed Hussain and Prof. Dimitri Vvedensky are also acknowledged for discussions. This work was first presented at PCSI-49 in Santa Fe.


\bibliography{EBR2}{}
\bibliographystyle{apsrev4-2.bst}

\end{document}